\documentclass{siamltex}
\usepackage{mathrsfs}
\usepackage{amsfonts,amssymb}
\usepackage{amsmath}
\usepackage{graphicx}
\usepackage{epstopdf}
\usepackage{color}
\usepackage{array}
\usepackage{hyperref}
\usepackage{bm}

\newcommand{\PreserveBackslash}[1]{\let\temp=\\#1\let\\=\temp}
\newcolumntype{C}[1]{>{\PreserveBackslash\centering}p{#1}}
\newcolumntype{R}[1]{>{\PreserveBackslash\raggedleft}p{#1}}
\newcolumntype{L}[1]{>{\PreserveBackslash\raggedright}p{#1}}

\providecommand{\U}[1]{\protect\rule{.1in}{.1in}}
\newcommand{\ba}{\bm\alpha}
\newcommand{\bx}{\bm\xi}
\DeclareMathAlphabet{\mathsfsl}{OT1}{cmss}{m}{sl}

\newcommand{\dif}{\mathrm{d}}

\newcommand{\ve}{\varepsilon}
\newtheorem{remark}{Remark}

\numberwithin{equation}{section}

\numberwithin{equation}{section}
\newtheorem{algorithm}{Algorithm}
\numberwithin{table}{section}
\numberwithin{figure}{section}
\numberwithin{remark}{section}

\usepackage{graphicx}
\usepackage{subfigure}

\usepackage{etoolbox}
\mathchardef\mhyphen="2D
\DeclareSymbolFont{extraup}{U}{zavm}{m}{n}
\DeclareMathSymbol{\vardiamond}{\mathalpha}{extraup}{87}

\makeatletter

\newcommand{\Rmnum}[1]{\expandafter\@slowromancap\romannumeral #1@}
\makeatother
\newcommand{\revisions}[1]{{#1}}

\title{ Quantifying the influence of conformational uncertainty in biomolecular solvation
  \thanks{This work was supported by the U.S. Department of Energy, Office of Science, Office of Advanced 
Scientific Computing Research as part of the Collaboratory on Mathematics for Mesoscopic Modeling of Materials (CM4).
Pacific Northwest National Laboratory is operated by Battelle for the DOE under Contract DE-AC05-76RL01830.
We would like to thank Xiaoliang Wan, Wen Zhou and Tom Goddard for fruitful discussions.
H.~Lei acknowledges a travel grant from the IMA for a workshop on Uncertainty Quantification in Materials Modeling and a travel grant from the DOE for the Conference on Data Analysis 2014. } }
\author{H.~Lei\footnotemark[2], X.~Yang\footnotemark[2], 
B.~Zheng\footnotemark[2], G.~Lin\footnotemark[3],  
\ and
N.~A.~Baker\footnotemark[2]\ \footnotemark[4]}

\begin{document}

\maketitle

\renewcommand{\thefootnote}{\arabic{footnote}}
\renewcommand{\thefootnote}{\fnsymbol{footnote}}
\footnotetext[2]{Pacific Northwest National Laboratory, Richland, Washington WA 99352, USA}
\footnotetext[3]{Department of Mathematics, Purdue University, West Lafayette, IN 47906, USA}
\footnotetext[4]{email: Nathan.Baker@pnnl.gov}

\begin{abstract}
Biomolecules exhibit conformational fluctuations near equilibrium states, inducing uncertainty in various biological properties in a \textit{dynamic} way.
We have developed a general method to quantify the uncertainty of target properties induced by conformational fluctuations.
Using a generalized polynomial chaos (gPC) expansion, we construct a surrogate model of the target property with respect to varying conformational states. 
We also propose a method to increase the sparsity of the gPC expansion by defining a set of conformational ``active space'' random variables. 
With the increased sparsity, we employ the compressive sensing method to accurately construct the surrogate model.
We demonstrate the performance of the surrogate model by evaluating fluctuation-induced uncertainty in solvent-accessible surface area for the bovine trypsin inhibitor protein system and show that the new approach offers more accurate statistical information than standard Monte Carlo approaches.
\revisions{Furthermore}, the constructed surrogate model also enables us to \textit{directly} evaluate the target property under various conformational states, yielding a more accurate response surface than standard sparse grid collocation methods. 
In particular, the new method provides higher accuracy in high-dimensional systems, such as biomolecules, where sparse grid performance is limited by the accuracy of the computed quantity of interest.
Our new framework is generalizable and can be used to investigate the uncertainty of a wide variety of target properties in biomolecular systems. 
\end{abstract}

\section{Introduction}
\label{sec:introduction}

Biomolecular structures are inherently uncertain due to thermal fluctuations and experimental limits in structural characterization.
At equilibrium, a biomolecule samples an ensemble of states governed by an energy landscape.
For a biomolecule with well-defined native structure at an energetic global minimum, these states are generally located in the neighborhood of the native structure.
While the native equilibrium structure of a biomolecule provides essential insight, it is also important to understand conformational fluctuations of biomolecular systems and their impact on molecular properties.
In particular, it is of great interest to accurately quantify the uncertainty in these properties caused by stochastic conformational fluctuations.

Molecular dynamics (MD) simulations offer a powerful tool for examining the influence of conformational uncertainty on biomolecular properties \cite{Dror2012Biomolecular, Adcock2006Molecular}.
Over the past few decades, this approach has made great progress in the development of accurate empirical force field as well as efficient simulation algorithms \cite{Ponder2003Force}.
However, despite these advances, MD is still a very computationally expensive simulation approach, particularly for large biomolecular complexes.
Moreover, the finite durations of MD simulations are plagued with uncertainty in calculated properties due to non-ergodic sampling. 
Many coarse-grained (CG) models and methods have been developed to facilitate molecular simulation at larger length scales and longer time scales.
One popular approach is the elastic network model (ENM), which involves a harmonic approximation of molecular energy landscape.
It has been observed that the low-frequency normal modes of a biomolecular system can be reproduced using a single-parameter Hookean potential between neighboring residues \cite{Tirion_PRL_1996,Hali_Bahar_1997,Tama_Sane_Protein_Eng_2001}.
In particular, by only modeling \revisions{interactions} between the neighboring $\alpha\mhyphen$carbon ($C_{\alpha}$), ENMs are able to predict structural fluctuations (e.g., Debye-Waller or B-factors) with surprising accuracy \cite{Hali_Bahar_1997,Ati_Bahar_BJ_2001}.

The simplified potentials used by CG models such as ENM allow us to examine structural fluctuations in a 
semi-analytical manner.
However, \revisions{there does not exist an \textit{analytical formula} that directly leads from the structural fluctuations to target biomolecular properties computed from the structure.
Instead, given a specific biomolecular conformation (e.g., one snapshot of biomolecule structure under fluctuation), we still need further numerical computation to obtain the target properties.}
This leads to an important practical question: \revisions{how do we utilize the stochastic information obtained from these models to efficiently quantify the uncertainty of the target property induced by the biomoleculular conformational (structural) fluctuation?
In many applications, a single native conformation of a molecule is used when computing a properties such as molecular volume and area \cite{Lee1971Interpretation, Richmond1984Solvent, Connolly1983Solventaccessible}, electrostatic and solvation properties \cite{Ren2012Biomolecular, Roux1999Implicit}, titration states \cite{Alexov2011Progress}, and other quantities.}
However, these quantities are all sensitive to the structure of the molecule and therefore subject to uncertainty induced by conformational fluctuations.
\revisions{Many studies neglect this uncertainty; those which attempt to assess it are forced to resort to time-consuming monte carlo sampling over the numerous biomolecular conformation states.}

In the present work, we address this issue by providing a general framework to quantify conformation-induced uncertainty on various biomolecular properties.
In particular, we construct a surrogate model of a target quantity in terms of the \revisions{molecular conformational states}.
\revisions{The constructed surrogate model enables us to efficiently evaluate the statistical information of the target property, e.g., probability density function.}
To the best of our knowledge, this is the first demonstration of how a target property response surface -- including property uncertainty -- can be directly evaluated from the biomolecular conformational distribution. 

To construct the surrogate model, we adopt the generalized polynomial chaos (gPC) \cite{GhanemS91, XiuK02} and formulate the target property as an expansion of a set of gPC basis functions determined by the specific conformation states, \revisions{where the gPC coefficients are determined by the values of the target properties on a number of sampling conformation states.}
Within this framework, \revisions{numerical quantification of the conformation-induced uncertainty is formulated as the following problem}: how can we accurately and efficiently construct the gPC based surrogate model of the target property using limited sampling points within the high-dimensional conformational space?
Several probabilistic collocation methods (PCM) such as ANOVA \cite{MaZ10, FooK10, ZhangCK12, YangCLK12} and sparse grid methods \cite{XiuH05, GanapathyZ07, FooWK08, NobileTW08, MaZ09} have been proposed to accurately construct gPC expansions by selecting specific collocation points for sampling.
However, there are two fundamental barriers when directly applying these approaches to high-dimensional biomolecular systems with hundreds to thousands of degrees of freedom in CG representations.
The first barrier is the required number of sampling points, which can be too large for any gPC approach beyond a linear approximation.
Moreover, empirical evidence indicates that sparse grid methods are often limited to dimensions less than $\sim 40$ (e.g., see \cite{petras_sg}). 
The second barrier is the presence of limited accuracy in the calculation of target properties -- even in the absence of structural uncertainty.  
For example, many calculations related to biomolecular solvation properties are subject to errors in the discretization and numerical solution of the associated partial differential equations \cite{Baker2005Biomolecular, Harris2013Influence}.
The error between the true \revisions{values} and the computed \revisions{values} of these target properties can lead to erroneous results due to inhomogeneous weight distribution over the sampling points, as illustrated in this paper.
To circumvent these difficulties, we adopt an alternative \revisions{noncollocation} method based on compressive sensing \cite{CandesT05, DonohoET06, BrucksteinDE09} which reduces the influence of the limited accuracy of the target property while \revisions{taking advantage of} the sparsity of the gPC expansion.
The compressive sensing method was initially proposed for signal processing and later applied to wide range of applications, including uncertainty quantification frameworks \cite{Li10, DoostanO11, YanGX12, YangK13}.
\section{Stochastic model}
\label{sec:model}

\revisions{In this section, we \revisions{briefly} introduce a semi-analytical stochastic model based on the elastic network model presented in \cite{Tirion_PRL_1996, Hali_Bahar_1997,Tama_Sane_Protein_Eng_2001}.
The resulting harmonic system yields a Gaussian probability distribution for conformational states \cite{Ati_Bahar_BJ_2001} that is straightforward to use in stochastic models for uncertainty quantification.
In addition to the dimensionality reduction provided by the coarse-grained ENM, we note that further dimensionality reduction can be obtained for biomolecular target properties that have local dependence on structure; i.e., where the values associated with a particular property depend only on a subset of atoms in the molecule.}

\subsection{Full stochastic model of \revisions{conformational} fluctuation}
\label{sec:ANM}
We construct the stochastic conformation space of the \revisions{biomolecular} system based on the coarse-grained (CG) anisotropic network model (ANM) \cite{Ati_Bahar_BJ_2001}, a variant of the ENM where each amino acid residue is modeled as a single CG particle connected to neighboring residues by anisotropic harmonic potentials.
ANM can be viewed as a simplified CG model of normal mode analysis \cite{Go_Noguti_PNAS_1983, Brooks_Karplus_PNAS_1983, Levitt_Sander_JQC_1983,McCammon_Harvey_1987}, where the model potential does not rely on the complex atomic-detail force field.
Consider a \revisions{biomolecule} of $N$ residues, \revisions{we denote the 
$3N$-dimensional equilibrium position vector by $\overline{\mathbf{R}}^{T} = 
\left [\overline{\mathbf{r}}_{1}^{T} ~\overline{\mathbf{r}}_{2}^{T} 
\cdots~\overline{\mathbf{r}}_{N}^{T}\right ]$, where $\overline{\mathbf{r}}_{i}$ is 
a 3-dimensional vector representing the equilibrium position of residue $i$.  Similarly,
we denote the $3N$-dimensional instantaneous position vector 
$\mathbf{R}^{T} = \left [\mathbf{r}_{1}^{T} ~\mathbf{r}_{2}^{T} \cdots ~\mathbf{r}_{N}^{T}\right ]$, where $\mathbf{r}_{i}$ represents the instantaneous 
position vector of residue $i$. The fluctuation vector can then be defined by
$\Delta \mathbf{R} = \mathbf{R} - \overline{\mathbf{R}}$.} 
The harmonic approximation for the potential energy $V$ with respect to the 
instantaneous position $\mathbf{R}$ is given by
\begin{equation}
\revisions{
	V({\mathbf{R}}) = \frac{\gamma}{2} \sum_{i<j} (r_{ij} - \overline{r}_{ij})^2  
	h(r_c - \overline{r}_{ij}),
	\label{eq:ANM_V}}
\end{equation}
where $\overline{r}_{ij}$ and $r_{ij}$ represent the equilibrium and instantaneous 
distances between residue $i$ and $j$, $\gamma$ is a model parameter representing the elastic coefficient of the harmonic potential, $r_c$ is \revisions{the} cut-off distance of the harmonic potential, and $h$ is the Heaviside function.

Given the potential defined by Eq.~\eqref{eq:ANM_V}, the $3N \times 3N$ Hessian matrix has the form
\[ \mathbf{H} = \begin{pmatrix}
	& \mathbf{H}_{11}  & \mathbf{H}_{12}  \cdots  & \mathbf{H}_{1N}\\
	& \mathbf{H}_{21}  & \mathbf{H}_{22}  \cdots  & \mathbf{H}_{2N}\\
	& \vdots & \\
	& \mathbf{H}_{N1}  & \mathbf{H}_{N2}  \cdots  & \mathbf{H}_{NN}
\end{pmatrix} \]
with the element $\mathbf{H}_{ij}$ defined by
\[ \mathbf{H}_{ij} = \begin{pmatrix}
	& \partial^2 V/\partial X_{i} \partial X_{j} & \partial^2 V/\partial X_{i} \partial Y_{j} & \partial^2 V/\partial X_{i} \partial Z_{j}\\
	& \partial^2 V/\partial Y_{i} \partial X_{j} & \partial^2 V/\partial Y_{i} \partial Y_{j} & \partial^2 V/\partial Y_{i} \partial Z_{j}\\
	& \partial^2 V/\partial Z_{i} \partial X_{j} & \partial^2 V/\partial Z_{i} \partial Y_{j} & \partial^2 V/\partial Z_{i} \partial Z_{j}\\
\end{pmatrix}, \]
\revisions{where $X_i$, $Y_i$ and $Z_i$ represent the Cartesian coordinates of residues $i$.
We note that the rank of $\mathbf{H}$ is $3N - 6$ since $V$ is translationally and rotationally invariant.}
This harmonic form for the potential leads to Gaussian statistics for the conformational probability distribution \revisions{(e.g., individual residue position distribution)}. 
The \revisions{correlation between individual residue fluctuation} can be determined 
by the pseudo-inverse of the Hessian matrix $\mathbf{H}$ as \cite{Ati_Bahar_BJ_2001} 
\begin{equation}
	\mathbf{C} = \mathbb{E} \left[\Delta \mathbf{R} \Delta \mathbf{R}^{T}\right] = 
  \frac{k_B T}{\gamma}  \mathbf{H}^{-1},
	\label{eq:residue_fluc}
\end{equation}
where \revisions{$\mathbb{E}\left[\cdot\right]$ denotes the expectation}, 
$k_B$ is the Boltzmann constant, $T$ is the temperature.

We perform an eigendecomposition of $\mathbf{H}$
\begin{equation}
	\mathbf{H} = \mathbf{W} \mathbf{\Lambda} \mathbf{W} ^{T},  ~~~\mathbf{\Lambda} = {\rm diag}(\lambda_1, \cdots , \lambda_{3N-6}),
\end{equation}
where $\lambda_i$ is the $i\mhyphen th$ nonzero eigenvalue of $\mathbf{H}$.
$\mathbf{W}$ is a $3N \times (3N-6)$ matrix defined by
\begin{equation}
	\mathbf{W} = \left [\mathbf{w}_1 \mathbf{w}_2 \cdots \mathbf{w}_{3N-6} \right ],
\end{equation}
where $\mathbf{w}_{i}$ is the corresponding $i\mhyphen th$ eigenvector of $\mathbf{H}$. 
\revisions{Then, the correlation matrix can be written as} 
\begin{equation}
	\label{eq:C_corr}
	\mathbf{C} = \frac{k_B T}{\gamma} \mathbf{W} \mathbf{\Lambda}^{-1} \revisions{\mathbf{W} ^{T}} = \mathbf{U} \mathbf{U}^{T},
\end{equation}
where $\mathbf{U} = \left(\frac{k_B T}{\gamma}\right)^{\frac{1}{2}} \mathbf{W} \mathbf{\Lambda}^{-\frac{1}{2}}$. The stochastic conformation space can be given by 
\begin{subequations}
	\label{eq:corr_matrix}
	\begin{equation}
		\mathbf{R}(\mathbf{\bx}) = \overline{\mathbf{R}} + \Delta \mathbf{R}(\bx), 
	\end{equation}
	\begin{equation}
		\Delta \mathbf{R}(\bx) = \mathbf{U} \bx,
	\end{equation}
\end{subequations}
where $\bx = (\xi_1, \xi_2, \cdots , \xi_{3N-6})$ is an independent and identically distributed (i.i.d.) Gaussian random vector.
Given a value of $\bx$, the corresponding coarse-grained conformation is 
fully determined by Eq.~\eqref{eq:corr_matrix}, allowing us to calculate 
target properties, denoted by $X(\bx)$. 

\subsection{Reduced-dimensionality stochastic model for conformational fluctuations}
\label{sec:reduce_ANM}
\revisions{The dimension of the stochastic conformation space constructed by Eq.~\eqref{eq:corr_matrix} is $3N-6$, which can still be high when $N$ is large.
However, we note that} some target properties of interest, particularly those related to a specific residue $p$, may have only local dependence on the neighboring residues' conformation rather than depending on all degrees of freedom in the biomolecule.
For example, the solvent accessible surface area (SASA) of a specific residue $p$ \revisions{only depends} on the positions of \revisions{itself} as well as its neighboring residues within a certain cutoff distance \revisions{$r_c^{\{p\}}$}.
Under such circumstances, the full position fluctuation correlation matrix $\mathbf{C}$ can be \revisions{replaced} by a $3N \times 3N$ matrix $\mathbf{C}^{\prime}$ of larger sparsity where the $(i, j)$ element (a $3 \times 3$ matrix) is given by
\begin{equation}
	\mathbf{C}^{\prime}_{ij} =  \mathbf{C}_{ij} h(\revisions{r_c^{\{p\}}} - r_{ip}) h (\revisions{r_c^{\{p\}}} - r_{jp}),
	\label{eq:reduce_corr_element}
\end{equation}
$r_{ip} = \vert \mathbf{r}_p - \mathbf{r}_i\vert$, $r_{jp} = \vert \mathbf{r}_p - \mathbf{r}_j\vert$, and \revisions{$r_c^{\{p\}}$} is a
cut-off distance of residue $p$ \revisions{such that} $X^{\{p\}}$ is independent 
of the residue $i$ \revisions{if $r_{ip} >  r_c^{\{p\}}$}.  

\begin{figure}
\centering
	\includegraphics*[scale=0.4]{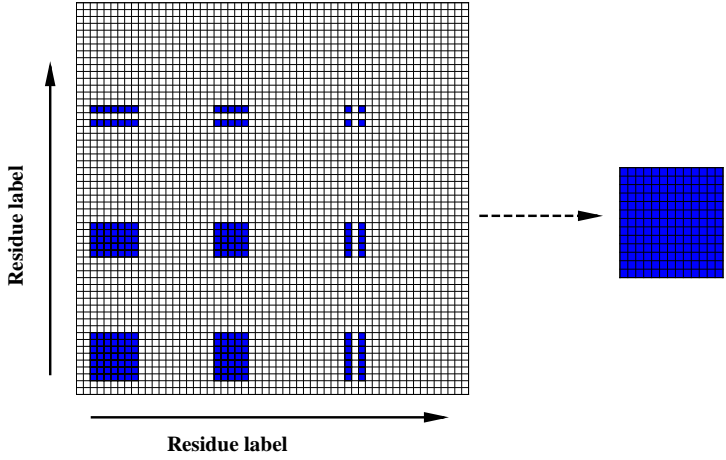}
	\caption{Sketch of a typical reduced property correlation matrix.
	The square on the left hand side represents the full correlation matrix.
	Each block represents $9$ elements (in $x$, $y$ and $z$ directions) of a residue in the correlation matrix $\mathbf{C}$. 
	The blocks in blue color represent the matrix elements associated with some \emph{local} target property $X$.
	The square on the right hand side represents the reduced correlation matrix $\mathbf{C}^{\{p\}}$ with lower dimensionality.}
\label{fig:corr_matrix}
\end{figure}

Fig.~\ref{fig:corr_matrix} illustrates this dimensionality reduction procedure for local properties as discussed above.
Similar to Eq.~\eqref{eq:corr_matrix}, we can construct the reduced stochastic conformation space by
\begin{subequations}
	\label{eq:reduce_corr_matrix}
	\begin{equation}
		\mathbf{C}^{\{p\}} = \mathbf{U}^{\{p\}} {\mathbf{U}^{\{p\}}}^{T}, 
	\end{equation}
	\begin{equation}
		\mathbf{R}^{\{p\}}(\bx^{\{p\}}) = \overline{\mathbf{R}^{\{p\}}} + \mathbf{U^{\{p\}}} \bx^{\{p\}},
	\end{equation}
\end{subequations}
where $ \bx^{\{p\}} $ is a $d$-dimensional i.i.d.\ normal random vector.

In summary, the value of a target property $X$ is determined by the specific conformation state, corresponding to a point $\bx$ (or $\bx^{\{p\}}$) in the full (or reduced) random space.
Our goal is to systematically quantify the uncertainty in $X$ with respect to the conformational fluctuations through gPC \revisions{expansion}, as 
introduced in \revisions{the} next section. 
\revisions{The rest of the manuscript focuses on local properties so we will} omit the superscript $\{p\}$ in \revisions{the following text}, \revisions{and use} $X(\bx)$ and $\bx$ to represent the target property and the $d$-dimensional random vector, respectively.
\section{Numerical methods}
\label{sec:method}
In this section, we first review the generalized polynomial chaos (gPC) expansion with a brief discussion on possible difficulties with \revisions{probabilistic} collocation methods.
Next, we introduce a \revisions{noncollocation} method to construct the gPC expansion based on compressive sensing.
\revisions{We note that the sparsity of gPC coefficients will affect the performance of compressive sensing method (see \cite{Candes08}).} 
Hence, we propose a method to elevate the sparsity of the gPC expansion by defining a new set of random variables according to the direction of variability in the target properties.

\subsection{gPC expansion and collocation method}
\label{sec:gPC}
We use the gPC expansion to construct the \revisions{surrogate model} of the target property $X$ with respect to the model parameter $\bx$, the molecular conformation by
\begin{subequations}
	\label{eq:gpc1}
	\begin{equation}
		{X}(\bx)=\sum_{|\ba|=0}^{\infty} c_{\ba}\psi_{\ba}(\bx), 
	\end{equation}
	\begin{equation}\label{eq:gpc_basis}
	\revisions{
\psi_{\bm\alpha}(\bm\xi)=\psi_{\alpha_1}(\xi_1)\psi_{\alpha_2}(\xi_2)\cdots\psi_{\alpha_d}(\xi_d), \quad \alpha_i \in \mathbb{N}\cup\{0\},} 
	\end{equation}
\end{subequations}
where $d$ is the number of random variables, $\ba=(\alpha_1,\alpha_2,\cdots,\alpha_d)$ is a multi-index, and $c_{\ba}$ is the
gPC coefficient to be determined.
\revisions{$\psi_{\alpha_i}(\xi_i)$ are univariate normalized Hermite polynomials, which satisfy the orthonormality condition:
\begin{equation}
	\int_{-\infty}^{\infty}\psi_k(\xi_i)\psi_l(\xi_i)\rho(\xi_i)\dif \xi_i = \delta_{kl},~~k, l \in \mathbb{N}\cup\{0\},
\end{equation}
where $\delta_{kl}$ is the Kronecker's delta and $\rho(\xi_i)=\dfrac{1}{\sqrt{2\pi}}\exp(-\xi_i^2/2)$ is the normal distribution function. 
The ENM described in Sec.~\ref{sec:ANM} relies on $d$ i.i.d.~standard normal random variables, hence the gPC basis functions are constructed
as the tensor products of univariate normalized Hermite polynomials as shown in Eq.~\eqref{eq:gpc_basis}.}

We truncate the expression \eqref{eq:gpc1} up to polynomial order $P$, hence $X$ is approximated as:
\begin{equation}\label{eq:gpc2}
	X(\bx)\approx\widetilde{X}(\bx)=\sum_{|\ba|=0}^{P}c_{\ba}\psi_{\ba}(\bx),
\end{equation}
using a total number of $n$ gPC terms \revisions{with $n = (P+d)!/(P! d!)$.}

Ideally, we would construct the truncated gPC expansion of $X(\bx)$ by computing $c_{\ba}$ using the \revisions{orthonormality} of $\psi_{\ba}$; e.g., \revisions{
\begin{equation}
	\label{eq:coef}
	c_{\alpha} = \int {X}(\bx)\psi_{\ba}(\bx)\rho(\bx)\dif\bx,
\end{equation}}
where \revisions{$\rho(\bx)$ is the probability density function (PDF) of $\bx$}. 

The integration can be accomplished by \revisions{utilizing} probabilistic collocation approaches such as tensor product \cite{Najm_UQ_2012_a, Najm_UQ_2012_b} or sparse grid \cite{XiuH05, GanapathyZ07, FooWK08} methods. 
\revisions{Specifically, by evaluating $X$ on} specific collocation points $\bx^1,\bx^2,\cdots,\bx^S$; \revisions{we have
\begin{equation}\label{eq:gpc_coef_quad}
	c_{\ba}=\int X(\bx)\psi_{\alpha}(\bx)\rho(\bx)\dif\bx\approx\sum_{i=1}^S X(\bx^i)\psi_{\ba}(\bx^i) w^i,
\end{equation}
where $w^i$ is the corresponding weight associated with collocation point $\bx^i$.}

However, for the high-dimensional biomolecular systems considered the present work, the required number of collocation  points \revisions{$S$} can be  computationally intractable.
For example, a small biomolecular system with a 27-dimensional reduced conformation random space would require \revisions{$S = 7.6\times10^{12}$} tensor product collocation points to construct a quadratic-order gPC expansion.
Standard sparse grid method based on Gaussian quadrature and Smolyak construction \revisions{reduces} this to $1513$ sampling points; however, the required number of sampling points is fixed for each order of the gPC approximation (e.g., the required number of sampling points for a $3$rd-order approximation is $27829$) which makes it difficult to incorporate adaptive sampling strategies.

Also we note that $X$ is generally \revisions{accompanied by numerical error}; e.g., 
\begin{equation}
	X(\bx) = \bar{X}(\bx) + \phi,
	\label{eq:x_error}
\end{equation}
where $\bar{X}(\bx)$ is the true value of the target property and $\phi$ represents 
the \revisions{numerical error}. \revisions{In this work, we require $\phi$ to satisfy} 
\begin{equation}
	\vert \phi \vert \ll \vert X \vert,
	\label{eq:err_small}
\end{equation}
\revisions{so we can systematically study the accuracy of the constructed surrogate model using different numbers of sampling data.
In general, the condition $\vert \phi \vert \ll \vert X \vert$ is not essential to the application of gPC expansion to construct the surrogate model. 
$\phi$ provides a lower bound of the numerical error of the surrogate model; e.g., we should not expect the error of the surrogate model to be less than the error accompanied by the sampling data.}

Moreover, as will be shown in Sec.~\ref{sec:results}, the aliasing error and the numerical error $\phi$ associated with $X$ may lead to poor approximation of $c_{\ba}$ by using probabilistic collocation method even if Eq.~\eqref{eq:err_small} is satisfied. 
To overcome the above difficulties, we compute the gPC expansion by applying compressive sensing as described in Sec.~\ref{sec:cs}.

\subsection{Compressive sensing method}
\label{sec:cs}
To construct the gPC expansion in Eq.~\eqref{eq:gpc2}, we compute $X(\bx)$ on $M$ sampling points \revisions{$(\bx^1, \bx^2, \cdots, \bx^M)$ }, \revisions{which are generated according to the distribution of random variables $\bx$.}
\revisions{In the present work, $\bx$ are $d$-dimensional i.i.d.\ standard normal random variables.}
We discretize Eq.~\eqref{eq:gpc2} as a linear system \revisions{
\[\begin{pmatrix}
	\psi_{\bm\alpha_1}(\bm\xi^1) & \psi_{\bm\alpha_2}(\bm\xi^1) & \cdots & \psi_{\bm\alpha_n}(\bm\xi^1)\\
	\psi_{\bm\alpha_1}(\bm\xi^2) & \psi_{\bm\alpha_2}(\bm\xi^2) & \cdots & \psi_{\bm\alpha_n}(\bm\xi^2)\\
	\vdots & \vdots &  & \vdots \\
	\psi_{\bm\alpha_1}(\bm\xi^M) & \psi_{\bm\alpha_2}(\bm\xi^M) & \cdots & \psi_{\bm\alpha_n}(\bm\xi^M)
\end{pmatrix}
\begin{pmatrix}
	c_{\bm\alpha_1} \\  c_{\bm\alpha_2} \\ \vdots \\ c_{\ba_n}
\end{pmatrix}=
\begin{pmatrix} 
	X(\bm\xi^1) \\   X(\bm\xi^2) \\ \vdots \\ X(\bx^M)
\end{pmatrix} + 
\begin{pmatrix} 
	\ve^1 \\   \ve^2 \\ \vdots \\ \ve^M
\end{pmatrix}
\] }
or equivalently,
\begin{equation}\label{eq:linear_sys}
	\mathbf{\Psi} \bm c = {\bm X} + \bm\ve,
\end{equation}
\revisions{where $\mathbf{\Psi}$ is the ``measurement matrix" with entries $\mathbf{\Psi}_{i,j}=\psi_{\ba_j}(\bx^i)$, $\bm c=(c_{\ba_1},c_{\ba_2},\cdots,c_{\ba_n})^T$ is the vector of the gPC coefficients, $\bm X=(X(\bx^1),X(\bx^2),\cdots,X(\bx^M))^T$ is the vector consists of the outputs and $\bm\ve=(\ve^1,\ve^2,\cdots,\ve^M)^T$ is related to the truncation error.}

\revisions{Notice that $\mathbf{\Psi}$ is an $M\times n$ matrix, and we are interested in the case when $M<n$ or even $M\ll n$.
It is presented in \cite{Candes08} that if $\mathbf{\Psi}$ satisfies the restricted isometry property (RIP), we can estimate $\bm c$ by solving the following optimization problem:
\begin{equation}
	\label{eq:comp_sensing2}
	(P_{1,\epsilon}):\qquad \arg\min_{\bm c^*}\Vert \bm c^*\Vert_1 \quad\text{subject to}\quad \Vert\mathbf{\Psi}\bm c^*-\bm X\Vert_2\leq\epsilon,
\end{equation}
where $\epsilon=\Vert\bm\varepsilon\Vert_2$. 
The upper bound of the error $\Vert \bm c-\bm c^*\Vert_2$ is decided by $\epsilon$ and the sparsity of $\bm c$:} \revisions{
\begin{equation}
	\label{eq:cs_err}
	\Vert\bm c-\bm c^*\Vert_2 \leq C_1\epsilon + C_2\dfrac{\Vert \bm c-\bm c_s\Vert_1}{\sqrt{s}},
\end{equation}
where $C_1,C_2$ are constants, $s$ is a positive integer, and $\bm c_s$ is $\bm c$ with all but the $s$-largest entries set to zeros.
For $\bm c$ in the present work, ``sparse" means small $\Vert \bm c - \bm c_s\Vert_1$ with $s$ being smaller (or much smaller) than the length of $\bm c$.}
The $(P_{1,\epsilon})$ optimization problem can be solved using classical convex optimization solvers (e.g., \verb|CVX| \cite{GrantB}), sparse recovery software packages (e.g., \verb|SPGL1| package \cite{BergF08}, $\ell_1$\verb|-MAGIC| \cite{l1magic}), or the split Bregman method 
\cite{goldstein2009split,yin2008bregman,cai2009linearized,cai2009convergence}.
In this paper, we use \verb|SPGL1|.

To solve Eq.~\eqref{eq:comp_sensing2}, we need the value of $\epsilon$, which is generally not known \textit{a priori}.
In this work, we estimate $\epsilon$ using a cross-validation method \revisions{\cite{DoostanO11, YangK13}}.
We first divide $M$ sampling data into two parts denoted by $M_r$  and $M_v$.
Second, $\bm c$ is computed with $M_r$ sample points with a chosen \revisions{series of} tolerance error $\epsilon_r$.
\revisions{Next, an optimized estimate $\hat{\epsilon}_r$ is determined 
such that $\Vert \mathbf{\Psi}_{v} {\bm c} - {\bm X}_{v}\Vert_2$ is minimized, where
$\mathbf{\Psi}_{v}$ and ${\bm X_v}$ represent the submatrix of $\mathbf{\Psi}$ 
and the subvector of ${\bm X}$ corresponding to validation portion of 
the sampling data.}
Finally, we repeat the above process for different replicas of the sample points and determine \revisions{the optimal} $\epsilon$ as $\epsilon=\sqrt{M/M_r}\hat\epsilon_r$.
In this work, we set $M_r = 2M/3$ and performed the cross-validation for three replications.
\revisions{We note that verifying the RIP for a given matrix is a NP-hard problem.
The aforementioned cross-validation procedure also serves as the verification for applying the $\ell_1$ minimization method to approximate $\bm c$ as it estimates the error of $\epsilon$.}

\subsection{Sparsity recovery via a ``renormalized active'' random space}
\label{sec:sparse}
The performance of the compressive sensing method introduced above is closely related to the ratio between the numbers of sampling points $M$ and basis functions \revisions{$n$}, as well as the sparsity of the linear system in Eq.~\eqref{eq:linear_sys}.
In general, accuracy improves with either larger $M/n$ ratios or sparser target vectors $\bm c$.
One way to increase $M/n$ \revisions{(for a given $M$)} 
is to reduce the dimension of stochastic space \revisions{hence $n$ is reduced.}
Unfortunately, for biomolecular systems, the dimension of the stochastic conformation space is determined by the structure of the molecule and is not always amenable to direct reduction.
Constantine et al.~\cite{Cons_Wang_SIAM_2014} have developed an alternative approach that \revisions{can be used to increase} sparsity by analysis of variability in the target properties.
For the target $X(\bx)$ with respect to PDF $\rho(\bx)$, we define gradient matrix $\mathbf{G}$ by \cite{Cons_Wang_SIAM_2014} 
\begin{equation}
	\mathbf{G} = \mathbb{E} \left[\nabla X(\bx) {\nabla X(\bx)}^T\right],
\end{equation}
where $\nabla X(\bx)$ is the gradient vector defined by 
\revisions{
$\nabla X(\bx) = \left(\frac{\partial X}{\partial \xi_1}, \frac{\partial X}{\partial \xi_2},
    \cdots \frac{\partial X}{\partial \xi_d}\right)^{T}.$
}
We conduct the eigendecomposition
\begin{subequations}
	\begin{equation}
		\mathbf{G} = \mathbf{Q} \mathbf{K} \mathbf{Q}^{T}, ~~~~~ 
		\mathbf{Q} = \left [{\mathbf q}_1 ~\mathbf{q}_2 \cdots ~\mathbf{q}_{d} \right ],
	\end{equation}
	\begin{equation}
		\mathbf{K} = {\rm diag}(k_1, \cdots , k_{d}), ~~~k_1 \ge \cdots \ge k_d \ge 0,
	\end{equation}
\end{subequations}
where $\mathbf{q}_i$ is the $i\mhyphen$th eigenvector of $\mathbf{G}$.
Therefore, the target property $X$ exhibits the largest variability along the direction $\mathbf{q}_1$ while it exhibits the smallest variability along the direction $\mathbf{q}_d$.
This motivates the definition of a new random vector 
\begin{equation}
\revisions{
	{\bm \chi} = \mathbf{Q}^{T} \bx,
	\label{eq:chi_def}
}  
\end{equation}
where $\mathbf{Q}$ is unitary and \revisions{$\bm \chi = (\chi_1, \chi_2, \cdots, 
\chi_d)^T$ are i.i.d.~Gaussian variables since $\bm\xi$ are i.i.d. Gausian} (also similar to Ref.~\cite{Tip_Ghanem_2014}). Dependence of the target property $X$ on $\chi_i$ decreases from $\chi_1$ to $\chi_d$.
Therefore, if we represent $X$ by a gPC expansion with respect to $\bm \chi$, $X$ may depend primarily on the first few random variables. \revisions{Then} the gPC coefficients associated with other variables exhibiting much smaller value (or even close to 0), yielding \revisions{sparser $\bm c$} for the linear system defined in Eq.~\eqref{eq:linear_sys}. 
Hence, if we recover the gPC coefficients with respect to $\bm \chi$ in Eq.~\eqref{eq:linear_sys} \revisions{by} the compressive sensing method, we expect more accurate result than directly recovering \revisions{gPC coefficients with respect to} $\bx$.

Unfortunately, the gradient vector $\nabla X(\bx)$ is generally not known \textit{a priori}.
Direct evaluation of $\mathbb{E} \left[\nabla X(\bx) {\nabla X(\bx)}^T\right]$ is very computationally expensive: the cost of evaluation of $\nabla X(\bx)$ is proportional to the dimension of the $\bx$.
Therefore, we evaluate $\nabla X(\bx)$ by approximating it via the gPC expansion recovered from $\bx$; e.g.,
\begin{subequations}
	\label{eq:G_evaluate}
	\begin{equation}
		\mathbf{G} \approx \mathbb{E} \left[ \nabla X^{\rm gPC} (\bx) {\nabla X^{\rm gPC} (\bx)}^T\right]
	\end{equation}
	\begin{equation}
		X^{\rm gPC} (\bx) = \sum_{|\ba|=0}^{P}c_{\ba}^{\left\{\bx\right\}}\psi_{\ba}(\bx),
	\end{equation}
\end{subequations}
where the superscript $\left\{\bx\right\}$ represents gPC coefficients directly recovered from $\bx$.
Evaluation of $\mathbb{E} \left[ \nabla X^{\rm gPC} (\bx) \nabla X^{\rm gPC} (\bx)\right]$ is straightforward \revisions{ with respect to PDF $\rho(\bx)$}, can be used to define $\mathbf{Q}$ and, therefore, the new random basis ${\bm \chi} = \mathbf{Q}^{T} \bx$.
Finally, \revisions{new basis functions associated with new random variables $\bm\chi$} can be used to reconstruct the gPC expansion of $X$ with respect to the $\bm \chi$ which, in general, yields greater sparsity.

\begin{remark}
	We do not reduce the dimension of the conformational space in the above procedure. 
	Instead, we define a new basis spanning the random space based on the variability direction of the target property.
	This set of basis \revisions{functions} is \textit{not} universal, it depends on the specific target property $X$. 
\end{remark}

\begin{remark}
	The gradient matrix $\mathbf{G}$ is approximated by Eq.~\eqref{eq:G_evaluate}.
	Therefore, eigenvectors \revisions 
{$\left [\mathbf{q}_1 ~\mathbf{q}_2 \cdots ~\mathbf{q}_{d} \right ]$}
may not correspond exactly to the steepest decay directions of variability for the target property $X$.
	Nevertheless, we adopt Eq.~\eqref{eq:G_evaluate} to construct a ``rotated'' space that provides larger (if not optimal) sparsity.
\end{remark}

\revisions{\begin{remark}
	Notice that $\bm\xi$ are i.i.d.\ Gaussian random variables and so as $\bm\chi$ due to the matrix $\mathbf{Q}$ being unitary, the new basis functions associated with $\bm\chi$ are still tensor product of Hermite polynomials, i.e., of the same form as in Eq.~\eqref{eq:gpc1}.
\end{remark} }

We summarize the entire procedure presented above (Sections \ref{sec:model} and \ref{sec:method}) in Algorithm \ref{alg:omp_dpd}. 
In the next Section, \revisions{we} apply this framework to quantify uncertainty in biomolecular solvent accessible surface area properties in the 
presence of conformational fluctuations. 

\begin{algorithm}
	\label{alg:omp_dpd} [Procedure to construct the gPC response surface of a given target quantity $X$ with respect to a 
stochastic biomolecular conformation space.]
	\vskip 5pt
	Step 1. For a biomolecular system, \revisions{we model the potential energy using the harmonic elastic network approach so the conformation 
fluctuation is Gaussian-distributed.
	We construct the full stochastic conformation space given in Eq.~\eqref{eq:corr_matrix}.
	For ``local'' target properties, we further reduce the dimension of the stochastic conformation space as in Eq.~\eqref{eq:reduce_corr_matrix}. 
	We conduct eigenvalue decomposition of the correlation matrix and represent the fluctuation by a $d$-dimensional i.i.d.\ standard normal random vector denoted by $\bx$.}
    
	Step 2. Generate $M$ sampling points $\bx^1, \bx^2, \cdots, \bx^M$ based on the distribution of $\bx$.
	Numerically compute $X$ on $\bx^1, \bx^2, \cdots, \bx^M$ to obtain $M$ outputs $X^1, X^2, \cdots, X^M$ \revisions{(where $X^q = X(\bx^q)$)}.
Denote ${\bm X}= (X^1, X^2,\cdots, X^M)$ as the ``observation" in $(P_{1,\epsilon})$.
	The ``measurement matrix" $\mathbf{\Psi}$ is constructed as $\mathbf{\Psi}_{i,j}=\psi_{\ba_j}(\bx^i)$, where $\psi_{\ba_j}$ are the basis functions.
	The size of $\mathbf{\Psi}$ is $M\times n$, where $n$ is the total number of basis functions depending on $P$ in \eqref{eq:gpc2}.
		
	Step 3. Set the tolerance $\epsilon$ in $(P_{1,\epsilon})$ by employing the cross-validation method.
		
	Step 4. Solve the $\ell_1$ minimization problem \revisions{
	\begin{equation*}
		\arg\min_{\bm c^*}\Vert \bm c^*\Vert_1\quad\text{subject to}\quad \Vert\mathbf{\Psi}\bm c^*-\bm X\Vert_2\leq\epsilon.
	\end{equation*}}
	to obtain the gPC coefficients $\bm c^{\{\bx\}}$.
		
	Step 5. Evaluate the gradient matrix \revisions{$\mathbf{G}=\mathbb{E} \left[ \nabla X^{\rm gPC} (\bx) {\nabla X^{\rm gPC} (\bx)}^T \right ]$}, given $\bm c^{\{\bx\}}$ and define the random vector $\bm \chi$ by Eq.~\eqref{eq:chi_def}.
	\revisions{Compute the sample of $\bm\chi$ as $\bm\chi^q=\mathbf{Q}^{T}\bx^q, q=1,\cdots,M$}.

	Step 6. \revisions{Construct new ``measurement matrix" $\tilde{\mathbf{\Psi}}$ by setting $\tilde{\mathbf{\Psi}}_{ij}=\psi_{\ba_j}(\chi^i)$.
	Construct the gPC expansion of $X(\bm \chi)$ by repeating steps 3-4 on random vector $\bm \chi$ and using \revisions{$\bm X=(X^1, X^2,\cdots,
    X^M)^T$} that have been determined in step 2.}

\end{algorithm}
\section{Numerical Results}
\label{sec:results}

\begin{figure}
	\centering
	\includegraphics*[height=8cm]{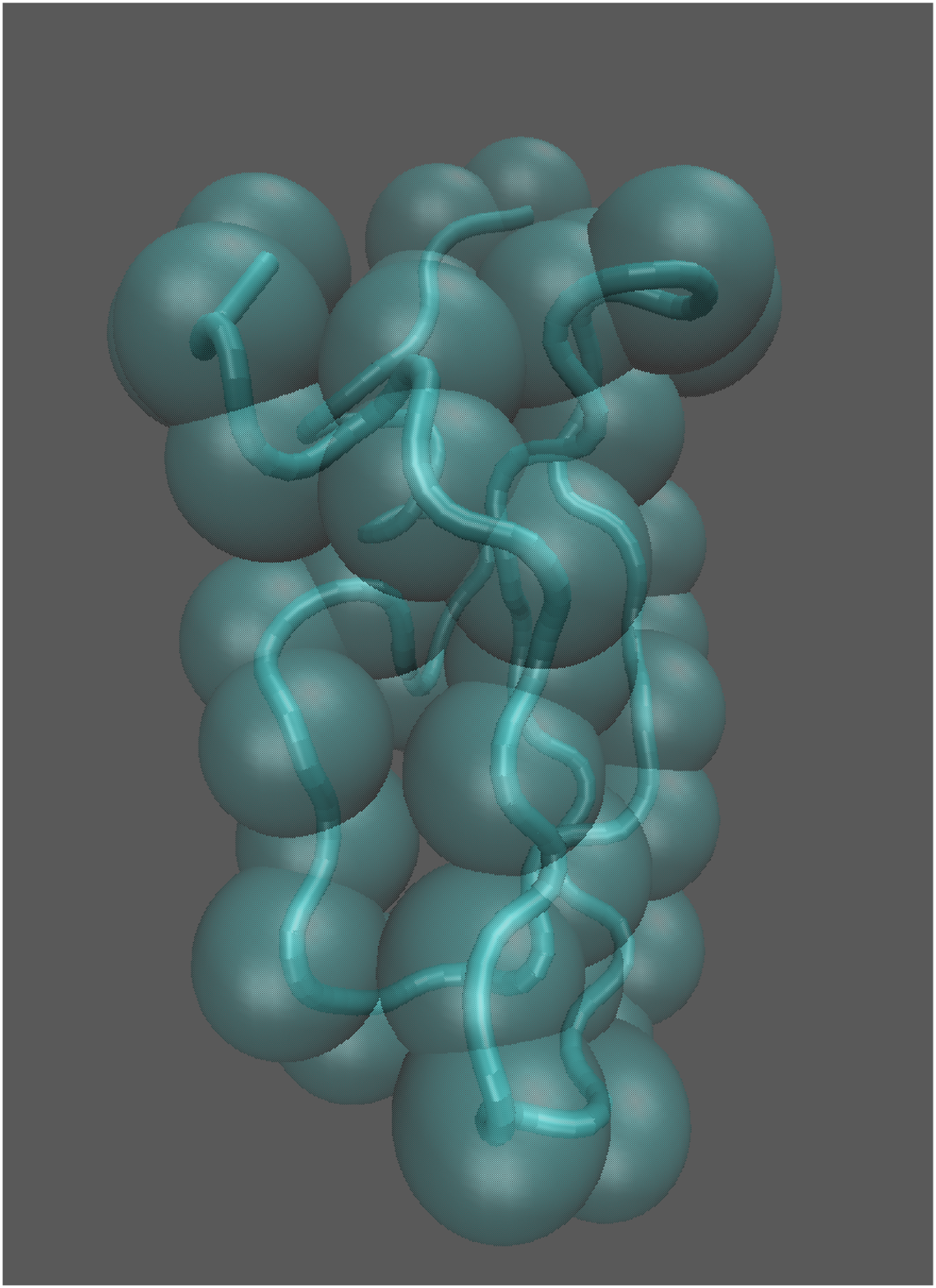}
	\includegraphics*[height=8cm]{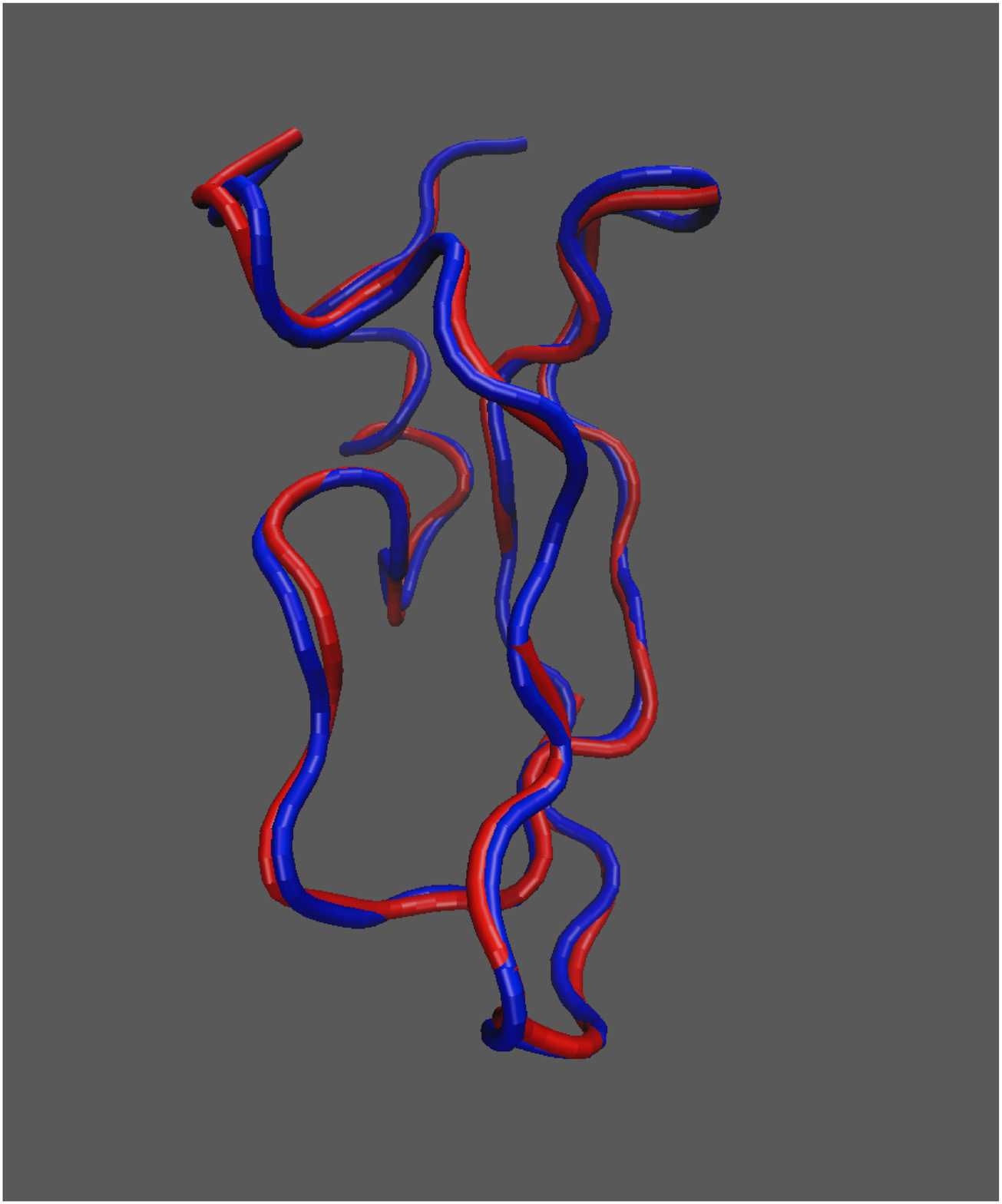}
	\caption{(a) Tube diagram of the equilibrium structure of bovine pancreatic trypsin inhibitor (PDB code: 5pti) with spheres denoting the residue $C_{\alpha}$ positions. 
	    (b) Tube diagrams of the molecule representing instantaneous conformational states under thermal fluctuation.}
	\label{fig:molecule_sketch}
\end{figure}

\revisions{As an example, we apply our method to quantify the uncertainty in solvent-accessible surface area (SASA) caused by conformational fluctuations in the biomolecule bovine pancreatic trypsin inhibitor (PDB code: 5pti) \cite{Wlodawer1984Structure}, shown in Fig.~\ref{fig:molecule_sketch}.}
\revisions{SASA is an essential element of numerous solvation models \cite{Baker2005Biomolecular, Roux1999Implicit, Ren2012Biomolecular}. 
The SASA for the entire molecule can be decomposed into residue-specific contributions, allowing us to explore the influence of conformational fluctuations on local area uncertainty.
SASA is calculated following Shrake et al.\ \cite{Shrake_JMB_1973}, setting $N^p$ nearly equidistant probing points on the solvent particle and determining SASA value for each residue from the fraction of probing points that are not buried by any of the neighboring residues.
In particular, we choose $N^p \approx 2.5\times10^5$ such that the numerical error $\phi$ satisfies $\left|\phi\right|/\left|X\right| \lesssim 1.0\times 10^{-4}$, see Sec.~\ref{subsec:err_analysis} for further discussion on sensitivity study on the accuracy of the constructed surrogate model by choosing different magnitudes of numerical error.}

To demonstrate the applicability of our method in exploiting information from limited sampling data, we focus on the performance of our method when constructing a surrogate model using less than 2500 sample data.
This performance is assessed relative to two reference systems:  a direct Monte Carlo simulation of the conformational space with $10^6$ sampling data as well as a system constructed by the standard sparse-grid collocation method.
We test our method by examining the $L_2$ error of the model as well as the Kullback-Leibler divergence between the probability density 
functions obtained from this new approach and the reference data.

\subsection{Surrogate model for SASA of individual residues}
\label{subsec:surrogate_model}
Fig.~\ref{fig:molecule_sketch} shows a sketch of the CG biomolecular model under equilibrium and thermal-fluctuation states.
Following Ref.~\cite{Ati_Bahar_BJ_2001}, each residue is modeled as a single $\alpha$-carbon particle as shown in Fig.~\ref{fig:molecule_sketch}(a).
Due to thermal fluctuations, the molecule exhibits a distribution of conformation states where individual residues may deviate from the equilibrium positions, as shown in Fig.~\ref{fig:molecule_sketch}(b).
To model the fluctuation of individual residues, we construct the ANM correlation matrix $\mathbf{C}$ by Eq.~\eqref{eq:C_corr} using a cut-off distance for the harmonic potential of $r_c = 9.8$\AA.
The radius values of the $\alpha$-carbon residue and the solvent probe were set to $2.8$ and $1.2$ \AA, respectively, for the SASA calculations.

\begin{figure}
	\centering
	\includegraphics*[trim = 0mm -20mm 0mm 0mm, clip, scale=0.42]{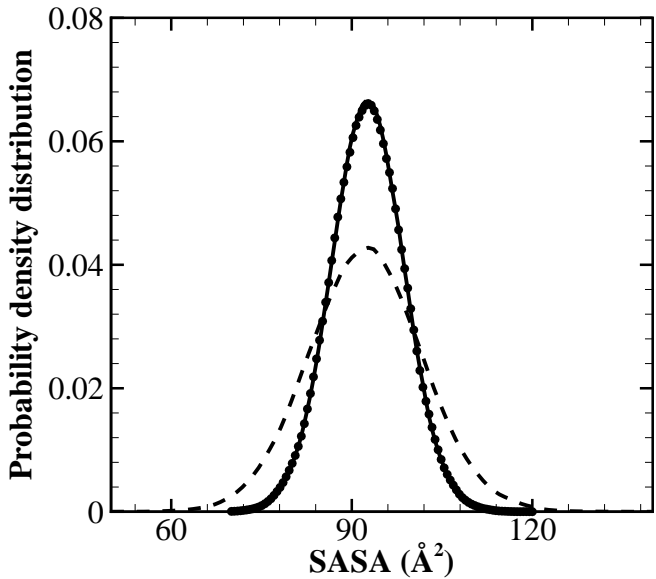}
	\caption{Probability density function of the SASA of the 14th residue obtained from the full correlation matrix $\mathbf{C}$ (solid line) and the local reduced correlation matrix $\mathbf{C}^{\prime}$ (``$\bullet$'' symbol).
	The dashed line represents the distribution obtained from the reduced correlation matrix where off-diagonal elements are set to zero.}
	\label{fig:pdf_validation}
\end{figure}

We first consider local properties and study the SASA of residue P$14$.
Starting with the full $168$-dimensional random correlation matrix $\mathbf{C}$, we construct the local correlation matrix $\mathbf{C}^{\prime}$ via Eq.~\eqref{eq:reduce_corr_element} by setting the neighbor cut-off distance $r_c^p$ to be $9.5$ \AA.
This cutoff value yields $8$ neighboring residues and therefore a $27-$dimensional random space $\mathbb{R}^{27}(\bx)$ by Eq.~\eqref{eq:reduce_corr_matrix}. 
As shown in Fig.~\ref{fig:pdf_validation}, the PDFs of the SASA of residue P14 extracted from the local and the full random conformation spaces agree well with each other, indicating that this particular property can be represented within a reduced space rather than the full $168$-dimensional space.
The dashed line in Fig.~\ref{fig:pdf_validation} represents the PDF extracted from the local random space by neglecting the fluctuation correlation between different residues (e.g., setting the off-diagonal blocks to zero).
The resulting distribution is wider than \revisions{that} predicted by the full 
correlation matrix. \revisions{This} is not surprising since the off-diagonal elements represent the harmonic potential contribution of molecular deformation in Eq.~\eqref{eq:ANM_V}.
Neglecting the off-diagonal block elements results in a more ``flexible'' molecule model which lacks the harmonic restraints and therefore exhibits a wider distribution of SASA values.

\begin{figure}
	\centering
	\includegraphics*[scale=0.4]{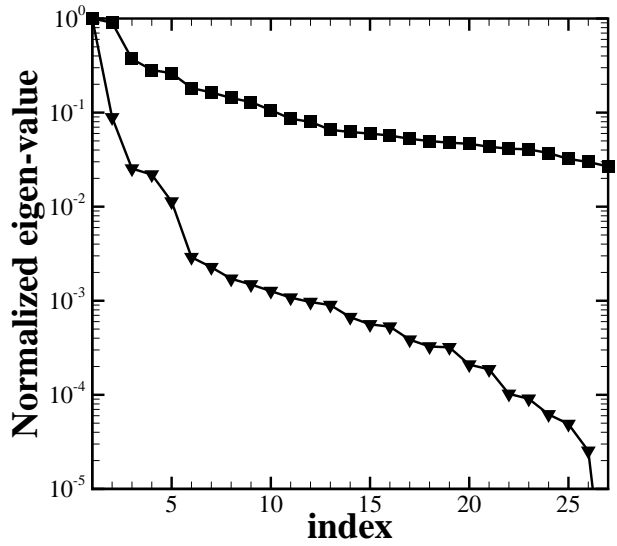}
	\caption{Normalized eigenvalues of the gradient matrix $\mathbf{G}$ (``${\large{\blacktriangledown}}$'' symbol) and the correlation matrix $\mathbf{C}^\prime$ (``${\blacksquare}$'' symbol). }
	\label{fig:eigen_graident_matrix}
\end{figure}

Next, we construct the surrogate model by computing the gPC coefficients within the reduced random space $\mathbb{R}^{27}(\bx)$ following the method presented in Sec.~\ref{sec:method}.
First, we calculate the gPC coefficients $c_{\ba}^{\{\bx\}}$ up to order  \revisions{$P=2$} ($406$ basis functions) by \revisions{setting $M=300$ in Algorithm $1$ and applying step $1-4$.}
Given $c_{\ba}^{\{\bx\}}$, we next construct the approximate gradient matrix $\mathbf{G}$ by Eq.~\eqref{eq:G_evaluate}.
Eigendecomposition of this matrix provides a set of rotated random variables ${\bm \chi}$ by Eq.~\eqref{eq:chi_def}, \revisions{(Step 5 in Algorithm $1$)}.                   
Fig.~\ref{fig:eigen_graident_matrix} shows the resulting normalized eigenvalues of $\mathbf{G}$ and the reduced correlation matrix $\mathbf{C}^{\prime}$. 
We note that $\mathbf{C}^{\prime}$ is independent of the target quantity $X$; it is completely determined by the molecular structure.
The eigenvalues of $\mathbf{C}^{\prime}$ decay slowly, at a rate similar to the full correlation matrix $\mathbf{C}$ (not shown in the plot), while the eigenvalues of the gradient matrix $\mathbf{G}$ decay much more quickly.
This result indicates that, for a particular quantity $X$, the eigenvectors of $\mathbf{C}$ do not necessarily correspond to the directions with the steepest decay of variability \revisions{in a target property}.

\begin{figure}
	\centering
	\includegraphics*[scale=0.4]{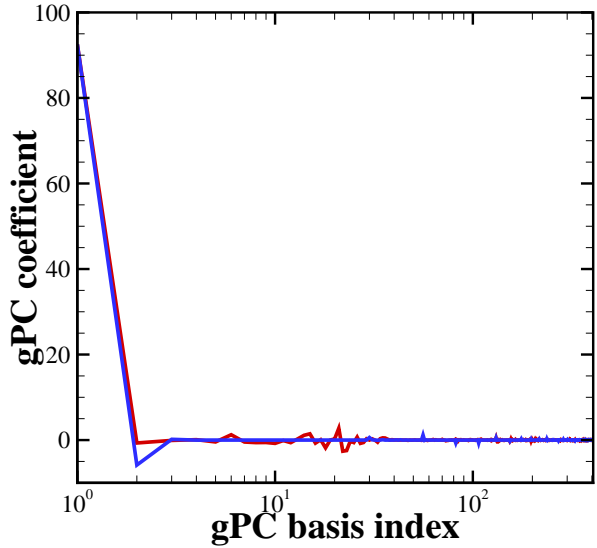}
	\caption{gPC coefficients (up to 2nd order) for the SASA value on the 14th residue obtained from CS method with respect to random vector $\xi$ (red) and $\chi$ (blue) with dimension $d=27$.}
	\label{fig:gPC_coeff_rotation}
\end{figure}

Given the variables ${\bm \chi}$, we compute the corresponding gPC coefficients $c_{\ba}^{\{\bm \chi\}}$ with order \revisions{$P=2$} \revisions{by applying Step 6\ in Algorithm $1$}.
The results are shown in Fig.~\ref{fig:gPC_coeff_rotation}.
Compared with $c_{\ba}^{\{\bm \chi\}}$, the spectrum of $c_{\ba}^{\{\bm \chi\}}$ exhibits a higher degree of sparsity, as expected.
This result indicates that, with the same polynomial order, the target quantity $X$ can be \revisions{approximated} using \textit{fewer} gPC terms with respect to the set of random variables $\bm \chi$ than with $\bm \xi$.

To examine the constructed surrogate model, we compute the relative $L_2$ 
error $\epsilon$ of the surrogate model by
\begin{equation}
\revisions{
	\label{eq:l2_err1}
	\epsilon = \left( \dfrac{\int |X(\bx) - \tilde X(\bx)|^2 \rho(\bx) \dif\bx} 
      {\int |X(\bx)|^2 \rho(\bx) \dif\bx} \right)^{1/2},}
\end{equation}
where $\tilde X$ is the gPC expansion of X by Eq.~\eqref{eq:gpc2} with $c_{\ba}^{\{\bx\}}$ and $c_{\ba}^{\{\bm \chi\}}$, respectively.
As $X(\bx)$ is unknown in general, we use Monte Carlo sampling to approximate the integral in Eq.~\eqref{eq:l2_err2}
\begin{equation}
\revisions{
	\label{eq:l2_err2}
	\epsilon \approx \left( \dfrac{\displaystyle \sum_{i=1}^{N_s} |X(\bx^i) 
      - \tilde X(\bx^i)|^2} {\displaystyle \sum_{i=1}^{N_s} |X(\bx^i)|^2} \right)^{1/2},}
\end{equation}
where $N_s$ is the number of sampling data.
In this work, we choose $N_s = 10^6$. 

\begin{figure}
	\centering
	\subfigure[]{
	\includegraphics*[scale=0.35]{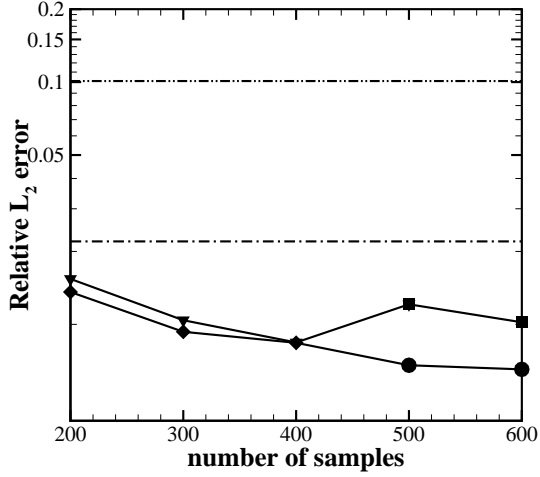}}
	\subfigure[]{
	\includegraphics*[scale=0.35]{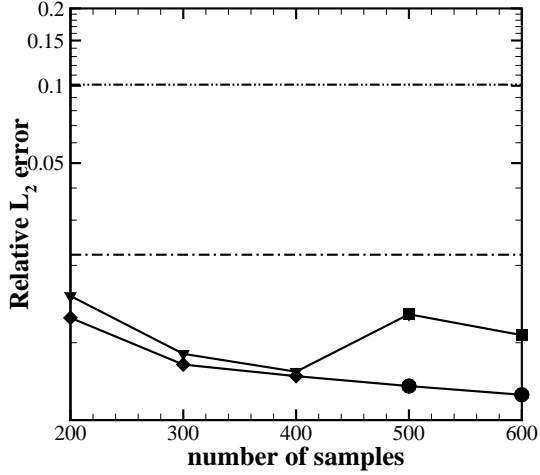}}
	\caption{Relative $L_2$ error of the SASA value on residue 14 predicted by the gPC expansions $\widetilde{X}({\bx})$ and $\widetilde{X}({\bm \chi})$, where the gPC coefficients are obtained from two separate sets of sampling data, represented by (a) and (b), respectively.
	The symbols ``${\large\large {\blacktriangledown}}$'' and ``${\blacksquare}$'' denote the $2nd$ and $3rd$ order gPC expansion by $\bx$.
	The symbols ``{$\vardiamond$}'' and ``{{\large $\bullet$}}'' denote the $2nd$ and $3rd$ order gPC expansion by $\bm \chi$.
	The dash-dot and dash-dot-dot lines represent the relative $L_2$ error of 1st and 2nd order gPC expansion obtained 
  from level-1 and level-2 sparse grid points, using $55$ and $1513$ sample points respectively.}
	\label{fig:L_2_residue_14}
\end{figure}

Fig.~\ref{fig:L_2_residue_14} shows the relative $L_2$ error of the constructed surrogate model with gPC coefficients recovered from two independent sets of sample data.
For each sample set, we use $200\mhyphen400$ points to construct the order $P = 2$ gPC expansion with $406$ basis functions and $500\mhyphen600$ sample points to construct the order $P = 3$ of gPC expansion with $4060$ basis functions.
For each case, the $L_2$ error decreases as we increase the number of sampling points from $200$ to $400$.
For the same number of sample points, the surrogate models constructed with respect to $\bm \chi$ exhibit smaller $L_2$ error than those constructed with respect to $\bx$.
In particular, given the same number of sampling points, sparser gPC coefficients $\mathbf{c}$ in Eq.~\eqref{eq:linear_sys} lead to more accurate recovery of $\mathbf{c}$ from the compressive sensing method by Eq.~\eqref{eq:comp_sensing2}.
\revisions{The accuracies of the compressive sensing methods based on $\bx$ 
and $\bm \chi$ are comparable} when the number of sampling points 
is close to the number of basis functions.

For random variables $\bx$, the $L_2$ error changes non-monotonically as we compute $c_{\ba}$ at order $P = 3$ by increasing numbers of sampling points.
The error increases as we increase the number of sample points to $500$ and then decreases as we increase the number of sample point to $600$ (although it remains larger than the $400$-point error).
This behavior is primarily due to the fact that the number of basis functions for $P = 3$ is much larger than the number of sample points and therefore $c_{\ba}^{\{\bx\}}$ is poorly recovered due to insufficient sample points.
However, for the transformed random variables $\bm \chi$, $c_{\ba}^{\{\bm \chi\}}$ can be accurately recovered due to the high sparsity of the gPC spectrum with a monotonic decrease in error with increasing numbers of sampling points.

We examined the surrogate model constructed by the sparse grid method based on Gaussian
quadrature collocation points \revisions{and Smolyak structure with gPC coefficients computed according to Eq.~\eqref{eq:gpc_coef_quad}}.
Fig.~\ref{fig:L_2_residue_14} shows the relative $L_2$ error of the surrogate model constructed by approximating the integral in Eq.~\eqref{eq:coef} with level-$1$ and level-$2$ sparse grid methods using $55$ and $1513$ sample points, respectively.
\revisions{Note that the algebraic accuracy of level-1 and level-2 sparse grid methods we use are $3$ and $5$, respectively.
Therefore, we construct $1$st-order and $2$nd-order gPC expansions with level-1 and level-2 methods respectively.}
The sparse grid results show systematically larger $L_2$ errors than the compressive sensing approach.
\revisions{An unexpected phenomenon is that the error of $2$nd-order expansion is larger than that of the $1$st-order expansion.  
This behavior will be explained in Sec.~\ref{subsec:err_analysis}.}

\begin{figure}
	\centering
	\subfigure[]{\includegraphics*[scale=0.4]{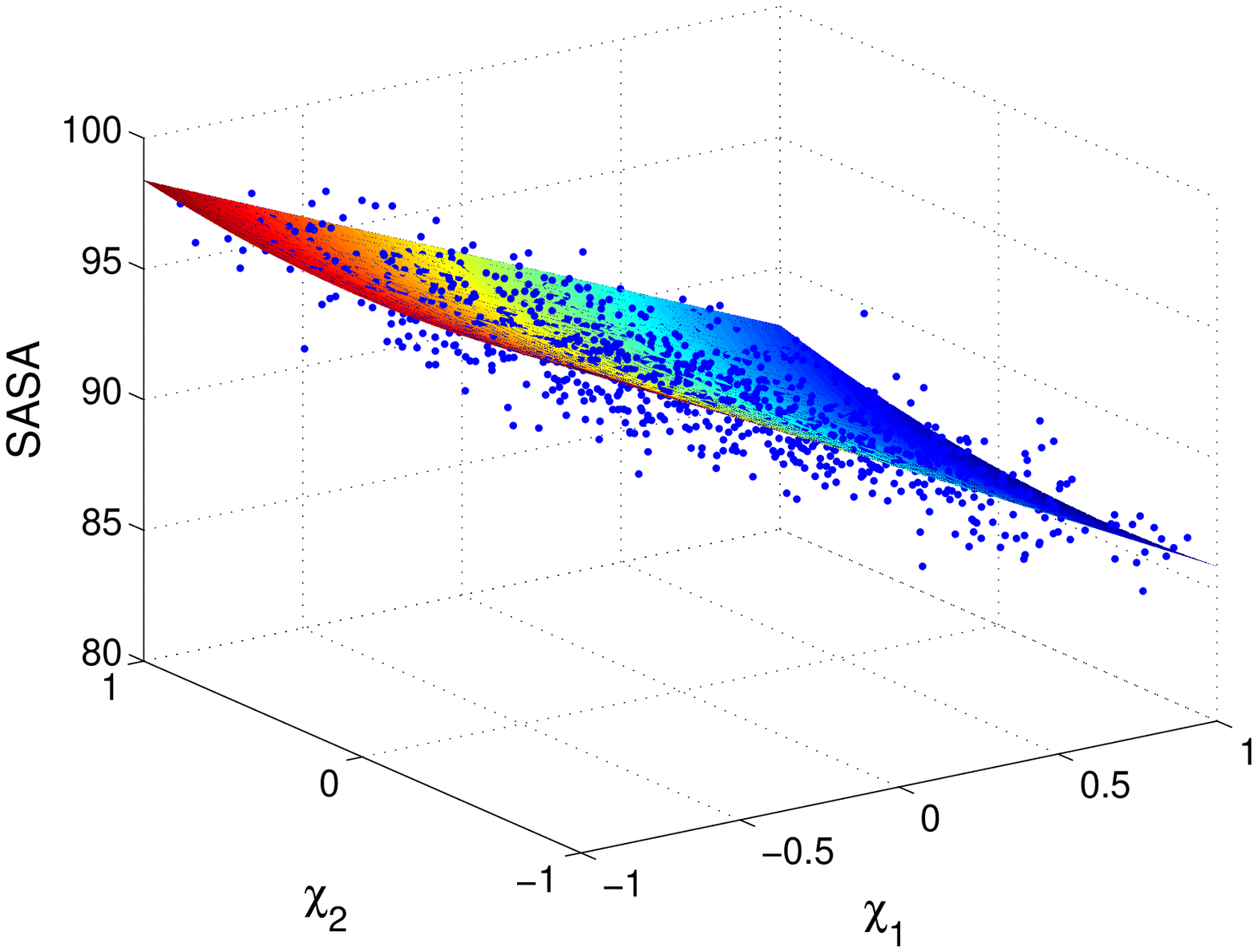}}
	\subfigure[]{\includegraphics*[scale=0.4]{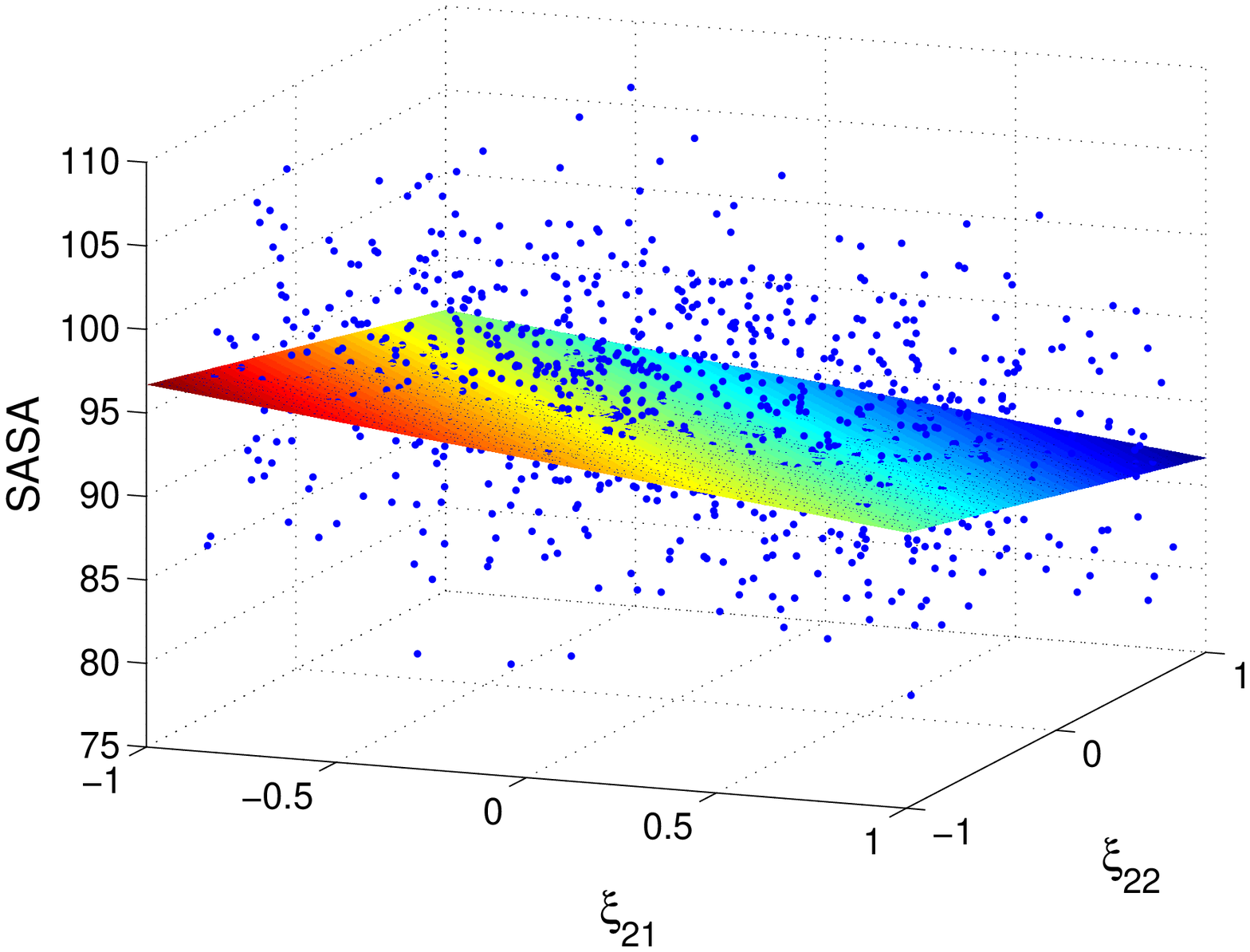}}
	\caption{(a) The \textit{reduced} response surface constructed by $\widetilde{X}(\chi_1, \chi_2, \chi_3^{0}, \cdots, \chi_{27}^{0})$, where $(\chi_3^{0}, \cdots, \chi_{27}^{0})$ are fixed values extracted from the i.i.d.\ normal distribution $\mathcal{N}(0, 1)$. 
	The scattered symbols (blue points) are direct numerical simulation results on stochastic points $(\chi_1, \chi_2, \cdots, \chi_{27})$ in $\mathbb{R}^{27}$ following an i.i.d.\ normal distribution $\mathcal{N}^{27}(0, 1)$. 
	(b) The \textit{reduced} response surface constructed by $\widetilde{X}(\xi_1^{0}, \cdots , \xi_{21}, \xi_{22}, \cdots , \xi_{27}^{0})$ where $(\xi_1^{0}, ..., \xi_{20}^{0}, \xi_{23}^{0}, ... , \xi_{27}^{0})$ are fixed values extracted from the i.i.d.\ normal distribution $\mathcal{N}(0, 1)$.
	The scattered symbols (blue points) are direct numerical simulation results on points $(\xi_1, \xi_{2}, \cdots, \xi_{27})$ following an i.i.d.\ normal distribution $\mathcal{N}^{27}(0, 1)$.}
	\label{fig:response_surface}
\end{figure}

The differences between the models constructed by $c_{\ba}^{\{\bx\}}$ and $c_{\ba}^{\{\bm \chi\}}$ can be further illustrated by examining the response surfaces in the \textit{reduced} random space shown in Fig.~\ref{fig:response_surface}.
This figure shows the response surfaces $\widetilde{X}(\bx)$ and $\widetilde{X}(\bm\chi)$ with respect to two random variables with the remaining $25$ random variables fixed.
The gPC coefficients are computed using $300$ sample points with the order $P = 2$ for both cases.
For $\widetilde{X}(\bm\chi)$, we only consider the first two random variables $\chi_1$ and $\chi_2$.
For $\widetilde{X}(\bx)$, we consider the random variables $\xi_{21}$ and $\xi_{22}$ which are associated with the largest magnitudes of the first order gPC coefficients.
For each case, we fixed the remaining variables as constant values extracted from an i.i.d.\ standard normal distribution $\mathcal{N}(0, 1)$.

The behavior of the Monte Carlo data around the response surfaces in Fig.~\ref{fig:response_surface} indicates that the variation of $\widetilde{X}(\bm\chi)$ strongly depends on $\chi_{1}, \chi_{2}$ while the dependence is much weaker for $\xi_{21}$ and $\xi_{22}$.
Furthermore, most of the symbols generated by $\widetilde{X}(\bm\chi)$ fall near the \textit{reduced} response surface $\widetilde{X}(\bm\chi)$ with small deviation while the deviations for $\widetilde{X}(\bx)$ are much larger around the response surface $\widetilde{X}(\bx)$.
As expected with rotation of the space by Eq.~\eqref{eq:chi_def}, this result indicates that $\widetilde{X}(\bm\chi)$ can be fitted fairly well by only using two variables.
However, if we use the original random variables, the reduced response surface can not be captured well even if we use the two most important variables associated with the first order gPC expansion. 
Fig.~\ref{fig:response_surface} clearly illustrates that the different sparsities of $\mathbf{c}$ result in different accuracies for the recovered response surfaces $\widetilde{X}(\bm\chi)$ and $\widetilde{X}(\bx)$.

\begin{figure}
	\centering
	\subfigure[]{\includegraphics*[scale=0.35]{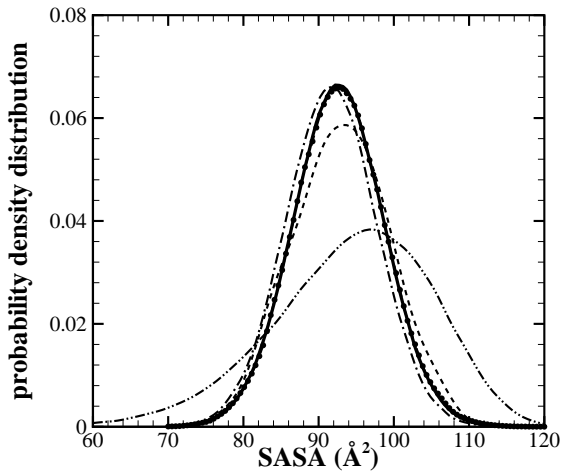}}
	\subfigure[]{\includegraphics*[scale=0.35]{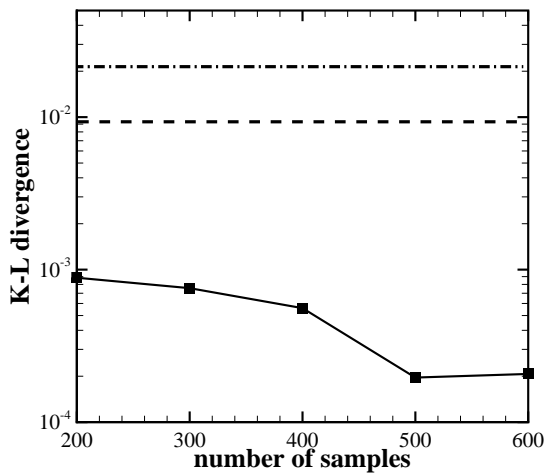}}
	\caption{(a) Probability density function (PDF) of the SASA values on residue P14 obtained from the gPC expansion $\widetilde{X}({\bm \chi})$ using $300$ sampling points (``$\bullet$'' symbol).
	Reference solution (solid line) is obtained from MC sampling using $10^6$ sample points.\revisions{
	Results of 1st and 2nd gPC expansion obtained from level-1 (dash-dot line, $55$ sample points) 
  and level-2 (dash-dot-dot line, $1513$ sample points) sparse grid method, and result from}, and
	direct MC sampling methods (dashed line, $300$ sample points) are also presented for comparison.
	(b) Kullback$\mhyphen$Leibler divergence between \revisions{the PDF of the reference solution} and the PDFs obtained from gPC expansion $\widetilde{X}({\bm \chi})$ (``${\blacksquare}$'' symbol) with varying numbers of sample points.
	Level-1 sparse grid (dash-dot line) and direct Monte Carlo (dashed line, $300$ sample points) results are presented for comparison.}
	\label{fig:Pdf_residue_14}
\end{figure}

To evaluate the statistical information extracted from the surrogate model, we compute the SASA PDF for target residue P$14$ by evaluating $10^6$ sampling data points with the constructed surrogate model.
These results are shown in Fig.~\ref{fig:Pdf_residue_14}(a) and compared with a reference solution based on the PDF computed from $10^6$ direct MC sample points.
The compressive sensing method with $300$ sample points yields the closest approximation of the reference solution.
In contrast, the PDFs constructed by the direct Monte Carlo and sparse grid methods show significant deviation from the reference solution.
To quantify the numerical error of the obtained PDFs, we computed the Kullback$\mhyphen$Leibler divergence 
\begin{equation}
	D_{\mathrm{KL}} = \int_{-\infty}^\infty \ln\left(\frac{f^{N}(X)}{f^{0}(X)}\right) f^N(X) \, {\rm d}X
\end{equation}
with the discrete form where $f^{N}(X)$ and $f^0(X)$ represent the PDFs of the numerical and reference solution, respectively.
For the compressive sensing method, $D_{\mathrm{KL}}$ decreases as we increase the number of sampling points, which is consistent with the $L_2$ error of the surrogate model (Fig.~\ref{fig:L_2_residue_14}).
The plateau value at $500\mhyphen600$ sampling points is primarily due to the finite resolution of the PDF:  a sensitivity study shows that $D_{\mathrm{KL}}$ between two i.i.d.\ sets of $10^6$ MC sample points is on the order of $10^{-4}$.
In contrast, $D_{\mathrm{KL}}$ values of the PDFs obtained by the level-1 and level-2 sparse grid methods are about $20$ and $450$ times larger (respectively) than the results of the compressive sensing method.

\subsection{Error sources and sensitivity analysis}
\label{subsec:err_analysis}
\begin{figure}
	\centering
	\subfigure[]{\includegraphics*[scale=0.27]{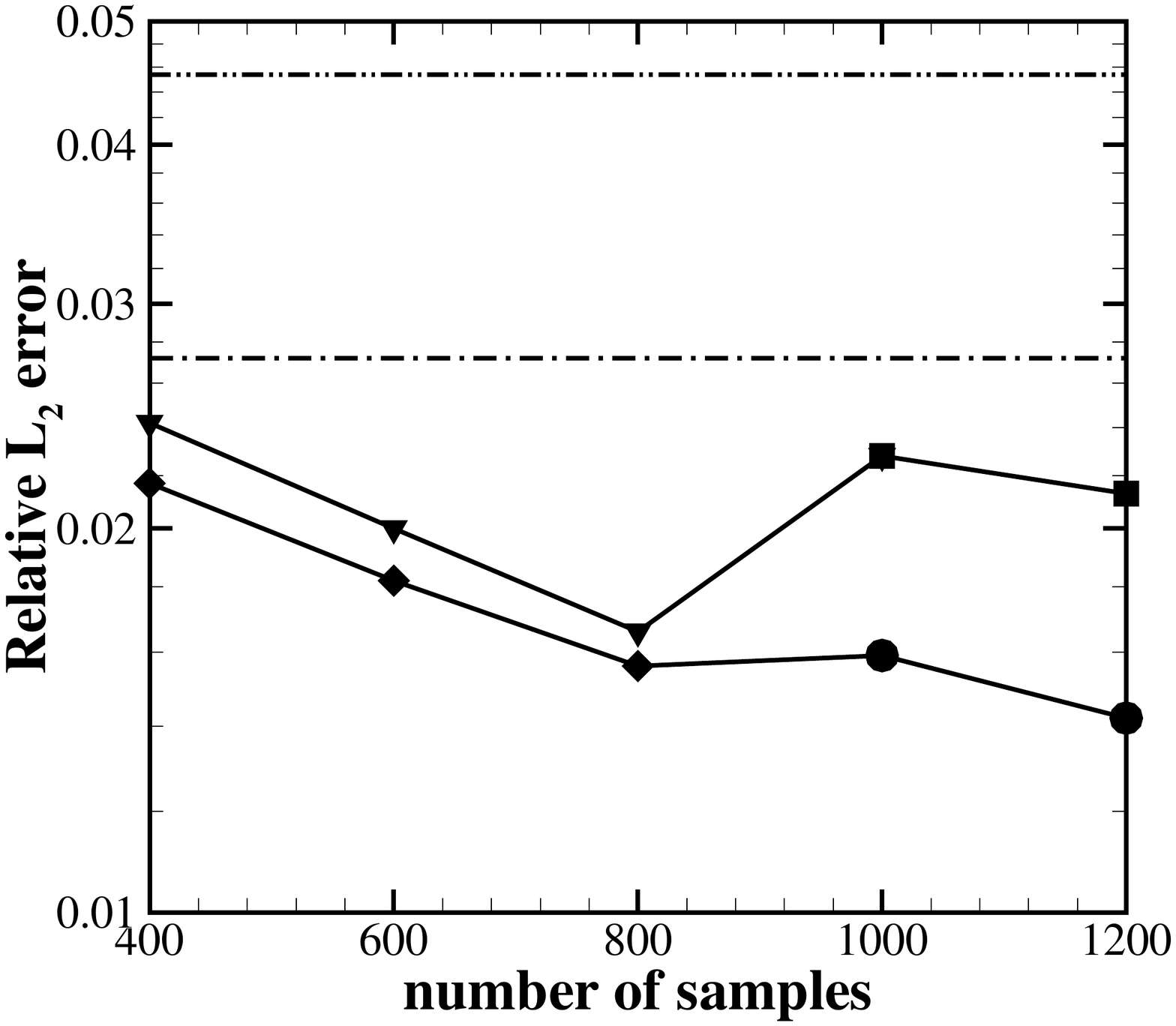}}
	\subfigure[]{\includegraphics*[scale=0.27]{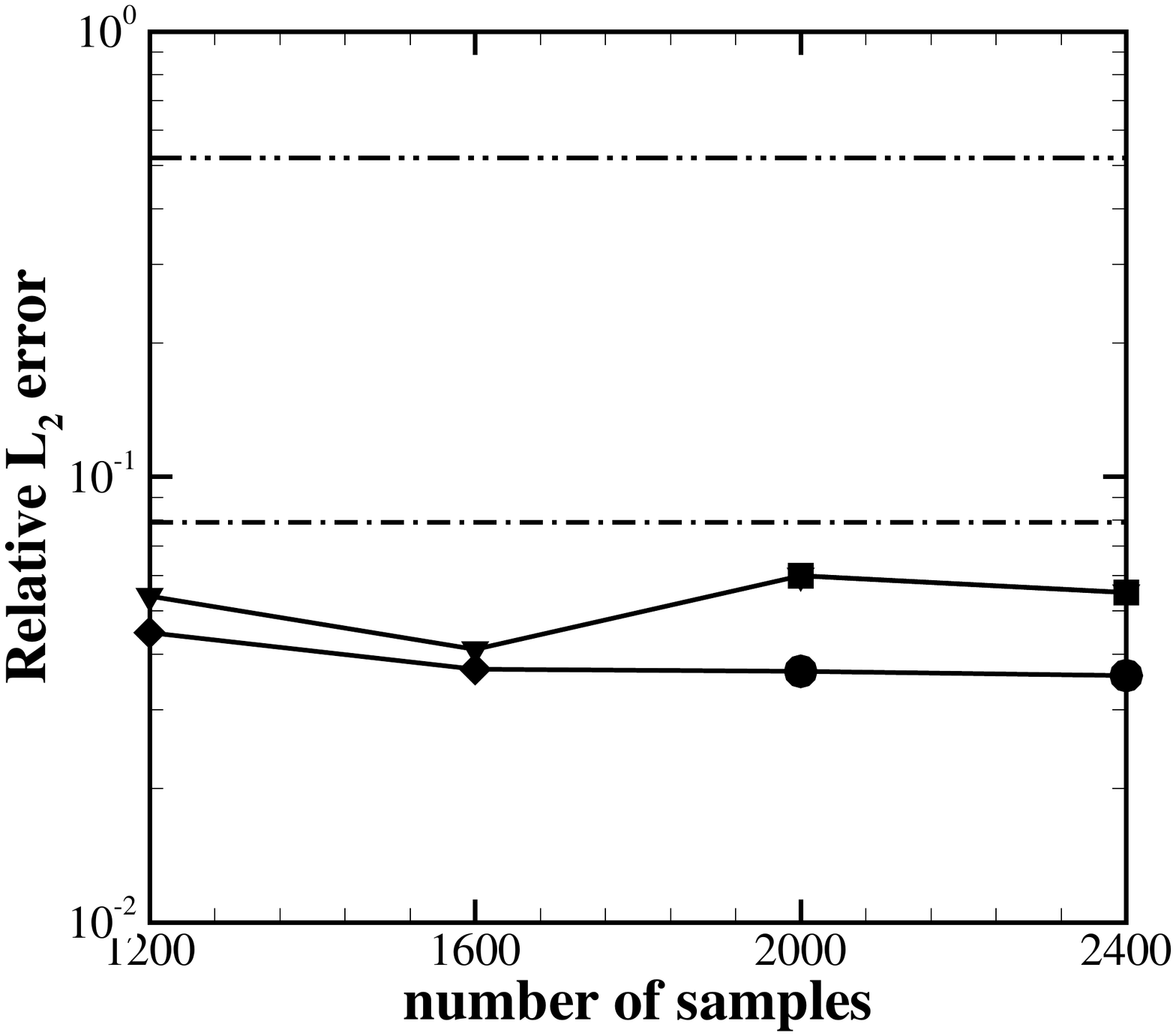}}\\
	\subfigure[]{\includegraphics*[scale=0.27]{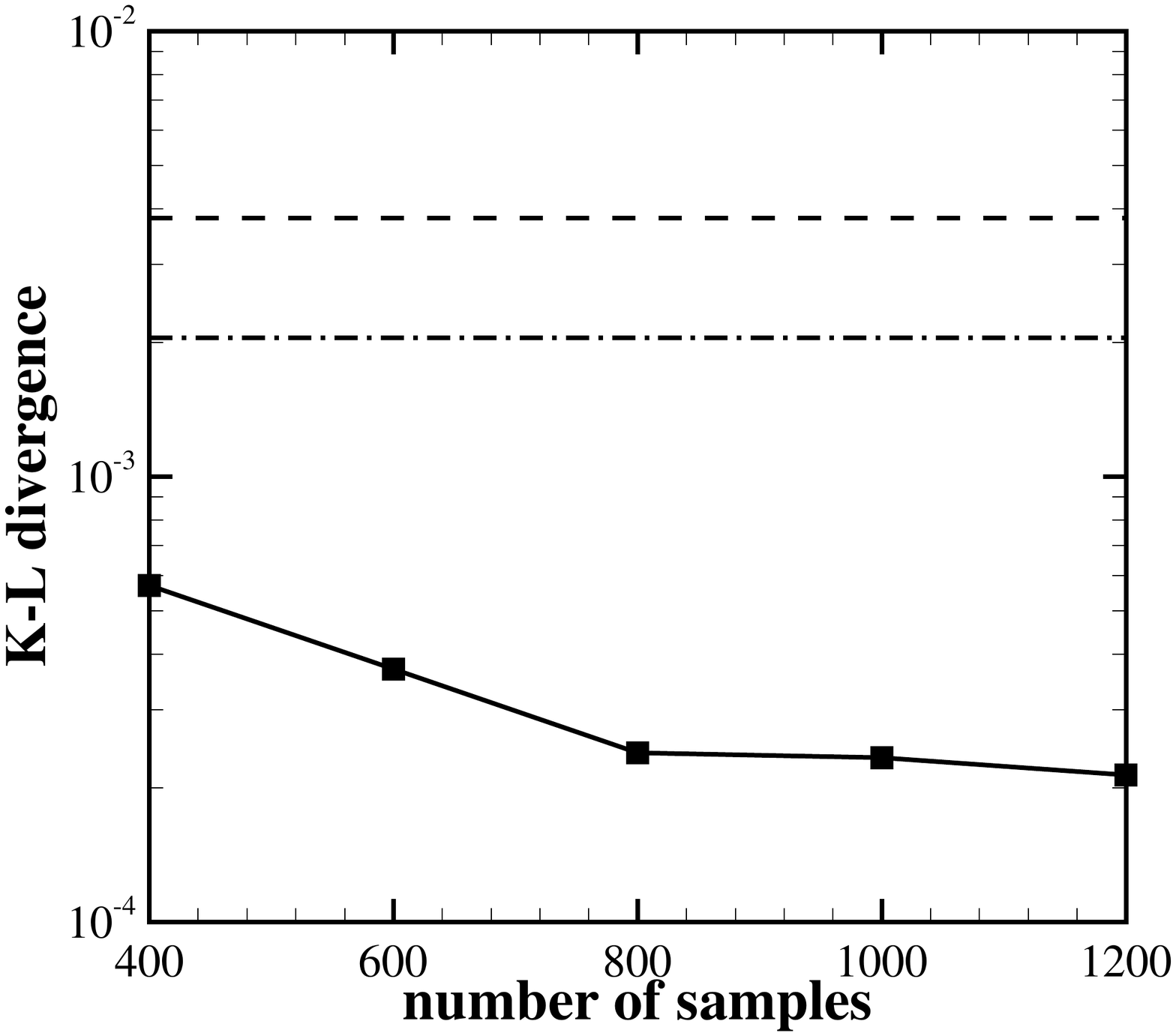}}
	\subfigure[]{\includegraphics*[scale=0.27]{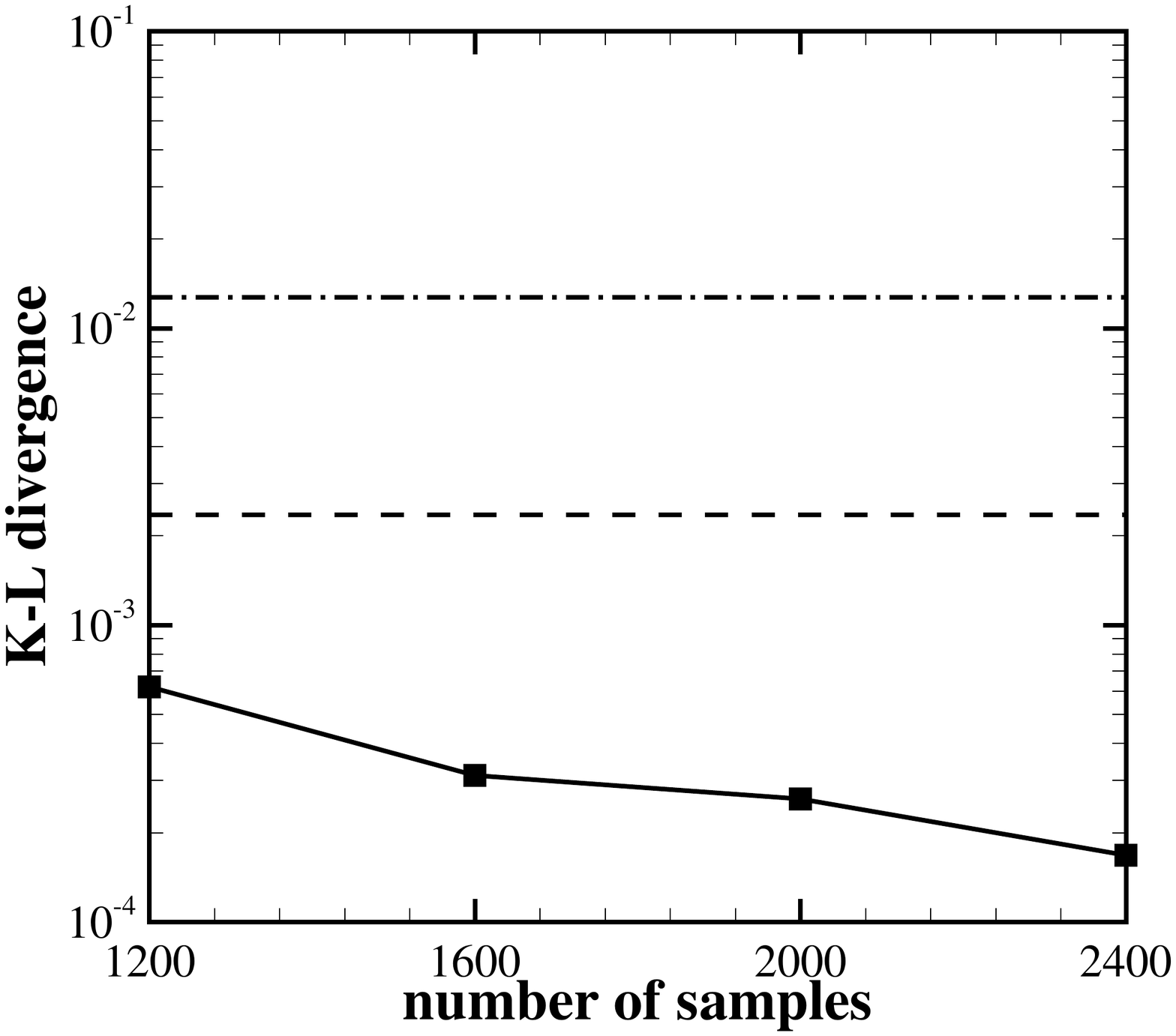}}
	\caption{(a-b) Relative $L_2$ error of the SASA value on residue 11 (a) and 20 (b) predicted by the gPC expansions $\widetilde{X}({\bm \xi})$ and $\widetilde{X}({\bm \chi})$.
    The symbols ``${\large {\blacktriangledown}}$'' and ``${\blacksquare}$'' denote the $2nd$ and $3rd$ order gPC expansion by $\bx$.
	The symbols ``{$\vardiamond$}'' and ``{{\large $\bullet$}}''  denote the $2nd$ and $3rd$ order gPC expansion by $\bm \chi$.
	The dash-dot lines represent the relative $L_2$ error \revisions{of 1st order gPC expansion} obtained from level-1 sparse grid methods using $85$ and $127$ sample points.
	The dash-dot-dot lines represent the relative $L_2$ error \revisions{of 2nd order gPC expansion} obtained from level-2 \revisions{sparse grid} methods using $3613$ and $8065$ sample points.
	(c-d) Kullback$\mhyphen$Leibler divergence between the \revisions{PDF of the reference solution} and the PDFs obtained from the constructed surrogate models (``${\blacksquare}$'' symbol) for residues 11 (c) and 20 (d).
	Level-1 sparse grid (dash-dot line) and direct Monte Carlo (dashed line, $300$ sample points) results are presented for comparison.}
	\label{fig:L_2_pdf_residue_11_20}
\end{figure}

To further investigate the applicability of the numerical methods for biomolecular systems, we quantified the SASA uncertainty for two other 
residues P$11$ and P$20$, which have $13$ and $20$ neighboring residues and correspond to random conformation spaces $\mathbb{R}^{42}$ and $\mathbb{R}^{63}$, respectively.
For each case, we constructed the surrogate model by the compressive sensing method with respect to both $\bx$ 
and $\bm \chi$, as well as by the level-$1$ (1st-order gPC expansion) and level-$2$ (2nd-order gPC expansion) sparse grid methods.
Fig.~\ref{fig:L_2_pdf_residue_11_20} shows the relative $L_2$ error of the surrogate models, the PDFs of the SASA values, and the $K\mhyphen L$ divergence with respect to the reference solution.

Similar to the results for residue P$14$, the surrogate models constructed with respect to $\bm \chi$ yield smaller error than the ones constructed with respect to $\bx$. 
The accuracy of the $\bx$ and $\bm \chi$ compressive sensing methods is comparable when the number of sampling points is close to the number of basis functions.
However, the surrogate model constructed with respect to $\bm \chi$ is more accurate than $\bm \xi$ when the number of sampling points is much less than the number of basis functions; e.g., when the third-order gPC terms are incorporated.
In particular, the surrogate model \revisions{($2$nd-order gPC expansion)} for residue $20$ constructed by the level-2 sparse grid in random space $\mathbb{R}^{63}$ yields the largest deviation from the reference solution.

\revisions{For the present system, the relatively large surrogate model error constructed by the sparse grid method (e.g., Eq.\ \eqref{eq:gpc_coef_quad}) can be explained as follows.
Given the target quantity $X$ computed at collocation points, the gPC coefficient $c_{\ba}$ is computed by }
\begin{equation}\revisions{
	\begin{aligned}
		c_{\ba}& = \sum_{i = 1}^{N^{\rm sp}} w^i(\bar{X}(\bx^i) + \phi (\bx^i))\psi_{\ba} (\bx^i)\\
		& = \sum_{i = 1}^{N^{\rm sp}} w^i(\bar{X}^i + \phi^i)\psi_{\ba}^i,
		\label{eq:c_sp}  
	\end{aligned}
}
\end{equation} 
\revisions{where $\bar X^i=\bar X(\bx^i), \phi^i=\phi(\bx^i), \psi_{\ba}^i=\psi_{\ba}(\bx^i)$ represent the true solutions, numerical error, and Hermite basis function evaluated at the sparse grid collocation point $\bx^i$, respectively; $N^{\rm sp}$ is the required number of sampling point with integral accuracy up to order $2P+1$; and $\phi$ is the associated numerical error accompanied with the computed value of $\bar{X}$.
We assume that} 
\begin{equation}
\revisions{
	\vert \phi^i \vert \ll \vert \bar{X}^i \vert
	\label{eq:error_small_2}}
\end{equation} 
\revisions{and that $c_{\ba}$ can be approximated by} 
\begin{equation}\revisions{
	\begin{split}
		c_{\ba} &= \sum_{i = 1}^{N^{\rm sp}} w^i\bar{X}^i\psi_{\ba}^i + \sum_{i = 1}^{N^{\rm sp}} w^i\phi^i\psi_{\ba}^i \\
		& = \bar{c}_{\ba}  + \sum_{\substack{\vert \bm\alpha + \bm\beta \vert \\> 2P+1}} \sum_{i=1}^{N^{\rm sp}} \bar{c}_{\bm \beta} w^i \psi_{\ba}^i \psi_{\bm\beta}^i + \sum_{i = 1}^{N^{\rm sp}} w^i \phi^i \psi_{\ba}^i,
	\end{split}
	\label{eq:sp_error}}
\end{equation}
\revisions{where $\bar{c}_{\ba} = \int \bar{X}(\bx) \psi_{\ba} (\bx)\rho(\bx) \dif\bx$ represents the true value of the gPC coefficient of index $\ba$ and $\bar{c}_{\bm \beta}$ represents the gPC coefficients with order $\vert \ba + \bm\beta \vert > 2P + 1$.
The second term on the righthand side of Eq.~\eqref{eq:sp_error} represents aliasing error due to the sparse grid approximation.
The third term $w^i\phi^i\psi_{\ba}^i$ represents the error due to the numerical error $\phi$ accompanied with the numerical computation of $\bar{X}$.} 

\revisions{For system of high dimensionality, both the aliasing error $\displaystyle \sum_{\substack{\vert \bm\alpha + \bm\beta \vert \\> 2P+1}} \sum_{i=1}^{N^{\rm sp}} \bar{c}_{\bm \beta} w^i \psi_{\ba}^i \psi_{\bm\beta}^i$ and the numerical error $\displaystyle \sum_{i = 1}^{N^{\rm sp}} w^i \phi^i \psi_{\ba}^i$ may induce pronounced error to the numerical computation of $c_{\ba}$.
Specifically, we assume that the numerical error $\phi_i$ superimposed on each collocation point is i.i.d.\ with zero mean and small variance $\vert \sigma_{\phi}^2 \vert \ll \vert X \vert^2$.
Given this assumption, the term $\sum_{i = 1}^{N^{\rm sp}} w^i \phi^i \psi_{\ba}^i$ is zero mean with variance 
\begin{equation}
	\mathrm{Var} \left(\sum_{i = 1}^{N^{\rm sp}} w^i \phi^i \psi_{\ba}^i\right) = \sum_{i = 1}^{N^{\rm sp}} (w^i)^2 (\psi_{\ba}^i)^2 \sigma_{\phi}^2.
    \label{eq:err_var}
\end{equation}}

\revisions{When the dimension of $\bx$ is large, we note that the weight distribution on sparse grid points is inhomogeneous; i.e., 
\begin{equation}
	\sum_i w^i = 1, ~~~~~~~\exists k,  \vert w^k \vert \gg 1.
\end{equation}
Fig.~\ref{fig:var_err_sp} plots the variance of the  term $\sum_{i = 1}^{N^{\rm sp}} \phi^i w^i $ (normalized by $\sigma_{\phi}^2$) for $c_{0}$
(i.e., \revisions{the coefficient of basis function $\psi_{\bm 0}(\bx)\equiv 1$} ) computed at different levels of sparse grid points. 
\begin{figure}
	\centering
	\includegraphics*[scale=0.4]{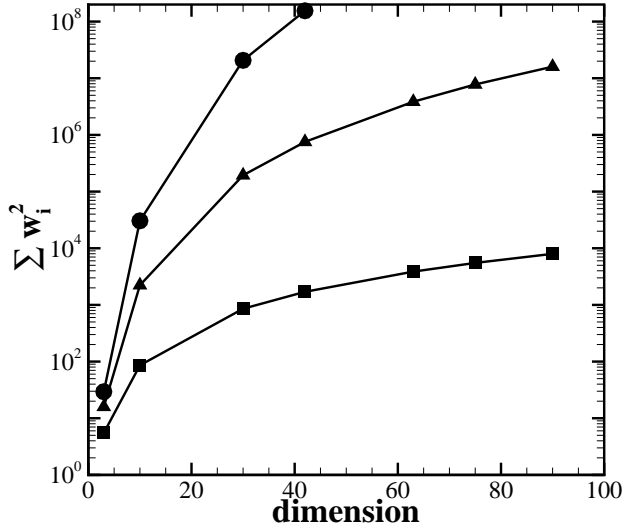}
	\caption{Variance of the numerical error term $\sum_{i = 1}^{N^{\rm sp}} \phi_i w_i $ (normalized by $\sigma_{\phi}^2$) for $c_{0}$ 
    computed by level-1 (``${\blacksquare}$''), level-2 (``{\large ${\blacktriangle}$}'') and level-3 (``{{\large $\bullet$}}'') sparse grid points.}
	\label{fig:var_err_sp}
\end{figure}
As the dimension increases, the variance of the error term increases rapidly, leading to non-negligible errors in the computation of $c_{\ba}$. }

\revisions{For illustration purposes, we consider the following 27-dimensional function:
\begin{equation}
	f(\bx) = 305+\sum_{k=1}^{27} \xi_k - 0.01\left(\sum_{k=1}^{27}\xi_k^2\right)^2.
\end{equation}
We first construct a 1st-order gPC expansion $f_1$ by using the level-1 sparse grid method (algebraic accuracy 3) to compute the coefficients:
\begin{equation}
	c_{\ba} = \int_{\mathbb{R}^{27}} f(\bx)\psi_{\ba}(\bx)\rho(\bx)\dif \bx \approx \sum_{i=1}^{N^{\rm sp}_1} f(\bx^i)\psi_{\ba} (\bx^i) w^i,
\end{equation}
where $\bx^i, w^i$ are sparse grid points and corresponding weights and $N^{\rm sp}_1$ 
  is the total number of the level-1 sparse grid points.
Next we construct a 2nd-order gPC expansion $f_2$ by using the level-2 sparse grid method (algebraic accuracy 5) to compute the coefficients with the same manner.
We compute the relative $L_2$ error of $f_1$ and $f_2$ as:
\begin{equation}
	\epsilon_k = \Vert f_k-f\Vert_2/\Vert f\Vert_2 = \left(\sum_{q=1}^{N^{\rm sp}_4}(f_k(\bx^q)-f(\bx^q))^2w^q\right)^{1/2}\Big / \left(\sum_{q=1}^{N^{\rm sp}_4}f(\bx^q)^2w^q\right)^{1/2},\quad k=1,2,
\end{equation}
where we use level-4 sparse grid method (algebraic accuracy 9) so that the 
numerical integral gives accurate result.
The 2nd-order expansion yields larger $L_2$ error due to the aliasing error in numerical integration.
In particular, $\epsilon_1 = 0.029$ while $\epsilon_2 = 0.100$.}

\revisions{We also constructed the gPC expansion of $f$ by sample points superimposed with numerical error:
\begin{equation}
	\label{eq:coef_pert}
	c_{\ba} = \int_{\mathbb{R}^{27}} f(\bx)\psi_i(\bx)\rho(\bx)\dif \bx \approx \sum_{q=1}^{N^{\rm sp}_1} f(\bx^q)(1+\sigma\zeta)\psi_i(\bx^q) w^q,
\end{equation}
where $\zeta$ is a standard Gaussian random variable and $\sigma$ is the magnitude of the noise.
We repeat each test with 1000 independent sets of noise and present the mean and standard deviation of the $L_2$ error in Table \ref{tab:perturb}.
As $\sigma$ increases from $10^{-4}$ to $10^{-3}$, the relative $L_2$ error further increases.
Moreover, the 2nd-order expansion yields much larger error than 1st-order expansion due to the more inhomogeneous weight distribution.
\begin{table}[!h]
	\centering
	\caption{$L_2$ error of of the 1st-order and 2nd-order expansion of $f$ constructed by level-1 ($\epsilon_1$) and level-2 ($\epsilon_2$) sparse grid method with sampling data superimposed with different magnitude of numerical error.}
	\begin{tabular} {C{5em}|C{16em}|C{16em}}
		\hline\hline
		$\sigma$ & $\epsilon_1$ & $\epsilon_2$ \\
		\hline
		$1\times 10^{-3}$ & $0.04  \pm 0.02 $ & $1.1 \pm 0.8$ \\
		$5\times 10^{-4}$ & $0.03  \pm 0.01 $ & $0.5 \pm 0.4$ \\
		$1\times 10^{-4}$ & $0.029 \pm 0.002 $ & $0.14 \pm 0.09$ \\
		\hline\hline
	\end{tabular}
\label{tab:perturb}
\end{table}
}

\revisions{Similar to the simple numerical example presented above, the surrogate model error of the biomolecular system constructed by the sparse grid method is determined by both the aliasing error and the numerical error on the sampling point.
Here we systematically investigate the $L_2$ error of the surrogate model of the target property $X$ (e.g., the SASA value) on residue P$14$.
The target quantity $X$ on sampling point is computed under various accuracy levels with relative error from approximately $10^{-5}$ to $10^{-3}$. 
The different accuracy levels are achieved by choosing different number of probe points on the solvent particle when computing the SASA value of the target residue.
For each accuracy level, we conduct a random 3D rotation of the molecule and conduct $32$ computation of the SASA value on residue P$14$.
We approximate the relative error by $\sigma_{X}/\mathbb{E}(X)$ with $\sigma_{X}$ and $\mathbb{E}(X)$ defined by
\begin{equation}
	\sigma_X = \frac{1}{S} \sum_{i=1}^{S} \sigma_{X^{i}}, \mathbb{E}(X) = \frac{1}{S} \sum_{i=1}^{S} X^{i},
\label{eq:variance_err}
\end{equation}
where $\sigma_{X^{i}}$ is the standard deviation of $32$ independent computation values of $X$ on sample point $\bx^{i}$ and $S$ is the total number of sample points.
We emphasize that $\sigma_{X}$ defined by Eq.~\eqref{eq:variance_err} is not equal to $\phi$ (e.g., the difference between the observation and true solution). 
However, $\sigma_{X}$ provides a useful guide to understand the magnitude of the disturbance on the sample data.
We also note that all the numerical results presented in Sec.~\ref{subsec:surrogate_model} were computed using sampling data with relative error 
$\sigma_{X}/\mathbb{E}(X) \approx 5\times 10^{-5}$}.

\revisions{ \begin{figure}
	\centering
	\includegraphics*[scale=0.4]{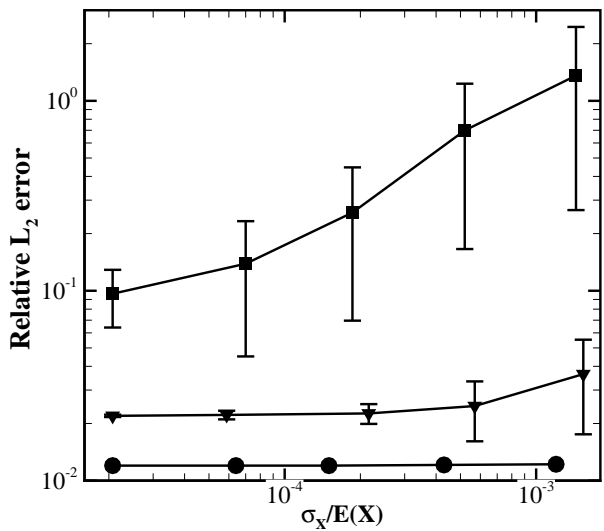}
	\caption{\revisions{Relative $L_2$ error of the SASA value on residue 14 predicted by 1st-order gPC expansion constructed by the level-1 sparse grid method (``${\blacktriangledown}$'' symbol), 2nd-order gPC expansion constructed by the level-2 sparse grid method (``${\blacksquare}$'' symbol), and 2nd-order gPC expansion constructed by compressive sensing (``{{\large $\bullet$}}'' symbol using $200$ sample points) under different accuracy levels $\sigma_{X}/\mathbb{E}(X)$. 
   For each accuracy level, $32$ sets of independent computations are conducted to compute the $L_2$ error of the constructed response surface.} }
   \label{fig:err_sp_perturb}
\end{figure}
}  

\revisions{Fig.~\ref{fig:err_sp_perturb} shows the relative $L_2$ error of the gPC expansion using the compressive sensing, level-1 sparse grid, and level-2 sparse grid methods.
The results from the sparse grid methods are very sensitive to accuracy level of the sample point.
For high accuracy levels, the $L_2$ error is mainly due to the aliasing error.
Increasing $\sigma_{X}/\mathbb{E}(X)$ from $2\times10^{-5}$ to $1.2\times10^{-3}$, the mean value of the relative $L_2$ error increases from $2.20\%$ to $3.36\%$ for the level-1 sparse grid method and from $9.66\%$ to $135.13\%$ for the level-2 sparse grid method.
In contrast, the compressive sensing method is insensitive to the imposed error on $X$ for the present system; the resulting error is nearly constant for $\delta \in \left[10^{-5}, 10^{-3}\right]$.
This result suggests another advantage of the present method: the present method is more stable in the presence of limited accuracy in the computed target quantity.
For high dimensional systems, the performance of sparse grid method strongly depends on the accuracy of the evaluation of $X$ at collocation points.
Similar phenomena have been reported previously \cite{ZhangTRK14}.
In practice, it may be computationally infeasible to evaluate $X$ at the accuracy level required for stable sparse grid results.
However, our new method based on compressive sensing shows a much weaker dependence on accuracy at individual sample points.}

\revisions{In order to explore the sensitivity of the accuracy level on the constructed gPC expansion, we conducted $32$ independent computations on the target quantity $X$ by randomly rotating the biomolecule $32$ times on each sampling point.
However, we are cautious to claim that numerical error $\phi^i$ superimposed on the $X^i$ is i.i.d.\ among the sampling points.
The i.i.d.\ assumptions adopted in Eqs.~\eqref{eq:err_var} and \eqref{eq:coef_pert} are used to demonstrate that the numerical error $\phi$ may further induce error to the constructed gPC expansion.
The study presented in this section demonstrates that the sparse grid method may induce relatively large errors to the constructed surrogate model. 
Rigorous error analysis of the sparse grid method in high-dimensional/complex systems is beyond the scope of this work.
However, there appear to be at least two important error sources (aliasing and numerical error on $X$) that could lead to erroneous results when applying the sparse grid method to high-dimensional systems such as biomolecules.}
We note that other specific structured or adaptive sparse grid methods \cite{Genz_Kei_1996, MaZ09, Nar_Xiu_Siam_2012, Nar_Xiu_Siam_2013} may alleviate the instability issue in high-dimensional systems. 
However, these methods either have less flexibility (the required number of sampling points is fixed for each accuracy level) or require a specialized design for adaptivity criteria. 

\subsection{Surrogate model for total molecular SASA}
\begin{figure}
	\centering
	\subfigure[]{\includegraphics*[scale=0.35]{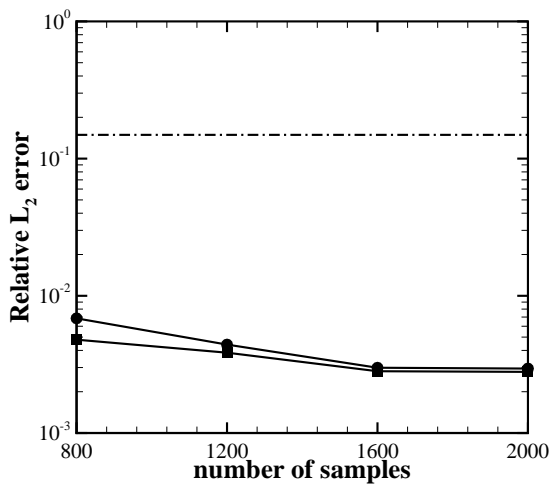}}
	\subfigure[]{\includegraphics*[scale=0.35]{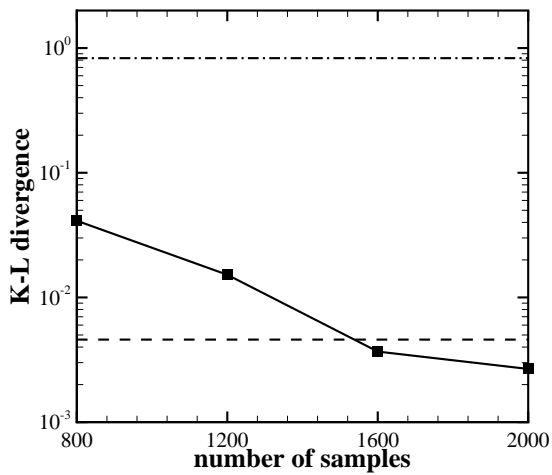}}
	\caption{(a) Relative $L_2$ error of the total molecular SASA by gPC expansion $\widetilde{X}({\bx})$ (``{{\large $\bullet$}}'') and $\widetilde{X}({\bm \chi})$ $\widetilde{X}({\bm \chi})$.
	The dash-dot line represents the relative $L_2$ error obtained from sampling on level-1 sparse grid points.
	Sampling over the level-2 sparse grid points generates erroneous results, as discussed in the text.  
	(b) Kullback$\mhyphen$Leibler divergence between the \revisions{PDF of the reference solution and the} PDFs obtained from surrogate models $\widetilde{X}({\bm \chi})$ (``${\blacksquare}$''), level-1 sparse grid (dash-dot line) and direct MD sampling (dash line, $2400$ sample points).}
	\label{fig:L_2_pdf_all_residue}
\end{figure}

Finally, we apply our method to quantify the uncertainty of the total SASA for the entire molecule.
Unlike the previous local per-residue SASA, this target quantity depends on the conformation states of all residues.
We construct the gPC expansion within the full random space $\mathbb{R}^{168}$. 
Due to the high dimensionality, we use a second-order gPC expansion with $14365$ basis functions.
Fig.~\ref{fig:L_2_pdf_all_residue} shows the relative $L_2$ error of the surrogate model and the K-L divergence of the PDFs.

\begin{figure}
	\centering
	\includegraphics*[scale=0.4]{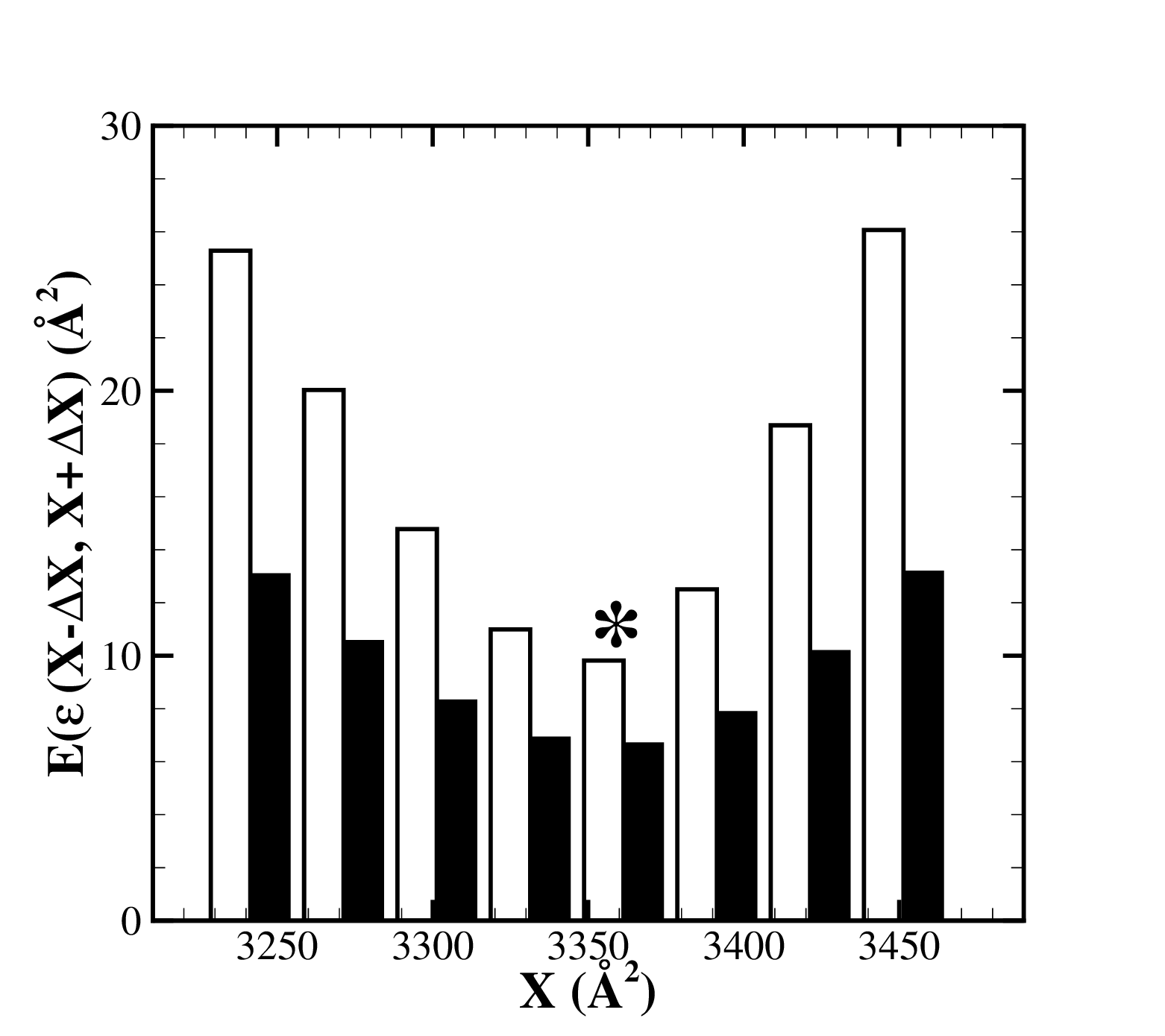}
	\caption{\revisions{The average error distribution of the surrogate models of 
   the total-SASA within different regimes $\left[X-\Delta X, X+\Delta X\right]$. 
   $\Delta X$ is chosen as $5$ \AA$^2$.}
	The surrogate models are constructed using 800 (blank) and 1600 (filled) sample points, respectively.
    The mean value of total SASA is approximately $3351$ \AA$^2$ (denoted by ``$\star$'' symbol), corresponding to conformations near the equilibrium state with respect to the thermal fluctuation.}
	\label{fig:err_dist}
\end{figure}

We note that the Hermite basis functions associated with the normal distribution are \textit{unbounded} which leads to inhomogeneous error distributions in the random space.
Fig.~\ref{fig:err_dist} shows the average error distribution of the surrogate model within different regimes of the SASA value.
The average error of the surrogate model of $X$ within $\left[x_1, x_2\right]$ is defined by
\begin{equation}
\revisions{
	\mathbb{E}\left(\epsilon(x_1, x_2)\right) = \left(\frac{\displaystyle \sum_i (X^{\rm gPC}(\bx^i) - X(\bx^i))^2 I_{(x_1, x_2)}(X(\bx^i))}
{\displaystyle \sum_i I_{(x_1, x_2)}(X(\bx^i))}\right)^{\frac{1}{2}},
}
\end{equation}
where \revisions{$I_{(x_1, x_2)}(X(\bx^i))$} is an indicator function which is 1 if \revisions{$X(\bx^i) \in \left [x_1, x_2 \right ]$} and 0 otherwise.
As shown in Fig.~\ref{fig:err_dist}, the error exhibits a minimum value near the equilibrium state and increases as the $X$ approaches the tails of the SASA PDF.
This demonstrates that the constructed surrogate model is not a \textit{global} approximation of the target quantity $X$ over the \textit{entire} random space.
Instead, it provides an approximation of $X$ with respect to the \textit{local} points near equilibrium within the random space.
Nevertheless, in practice, we are generally interested in the variation of $X$ with response to conformation fluctuation near the equilibrium state; e.g., the relatively small thermally induced molecular fluctuations considered in the present work. 

Similar to the ``local'' properties discussed above, the gPC expansion recovered by our compressive sensing method yields the smallest error.
However, the advantage of our new method over direct Monte Carlo sampling for the global SASA is not as large as in the case of local properties.
By constructing multi-D basis function through tensor product of one-dimensional basis functions, the upper bound (Here, the upper bound exists because the sampling of the Gaussian random variables is truncated in practice.) of the basis function becomes larger, which decreases the efficiency of the compressive sensing method.
This is similar to the phenomenon observed by others \cite{RauhutW12, YanGX12}.
If only statistical information such as expectation values or PDFs are needed, other methods such as quasi-Monte Carlo \cite{Nied92, Sloan94} may be suitable for high-dimensional systems.

\clearpage
\bibliographystyle{siam}
\bibliography{protein,uq}

\begin{thebibliography}{10}

\bibitem{l1magic}
{\em $\ell_1$-magic}.
\newblock \url{http://statweb.stanford.edu/~candes/l1magic/}.

\bibitem{Adcock2006Molecular}
{\sc S.~A. Adcock and J.~A. McCammon}, {\em Molecular dynamics:  survey of
  methods for simulating the activity of proteins}, Chem. Rev., 106 (2006),
  pp.~1589--1615.

\bibitem{Alexov2011Progress}
{\sc E.~Alexov, E.~L. Mehler, N.~A. Baker, A.~M. Baptista, Y.~Huang,
  F.~Milletti, J.~E. Nielsen, D.~Farrell, T.~Carstensen, M.~H.~M. Olsson, J.~K.
  Shen, J.~Warwicker, S.~Williams, and J.~M. Word}, {\em Progress in the
  prediction of {pKa} values in proteins}, Proteins, 79 (2011), pp.~3260--3275.

\bibitem{Ati_Bahar_BJ_2001}
{\sc A.~R. Atilgan, S.~R. Durell, R.~L. Jernigan, M.~C. Demirel, O.~Keskin, and
  I.~Bahar}, {\em Anisotropy of fluctuation dynamics of proteins with an
  elastic network model}, Biophys. J., 80 (2001), pp.~505--515.

\bibitem{Baker2005Biomolecular}
{\sc N.~A. Baker}, {\em Biomolecular applications of {Poisson}–{Boltzmann}
  methods}, Rev. Comp. Ch., 21 (2005), pp.~349--379.

\bibitem{Brooks_Karplus_PNAS_1983}
{\sc B.~Brooks and M.~Karplus}, {\em Harmonic dynamics of proteins: normal
  modes and fluctuations in bovine pancreatic trypsin inhibitor}, Proc. Natl.
  Acad. Sci. U.S.A., 80 (1983), pp.~6571--6575.

\bibitem{BrucksteinDE09}
{\sc A.~M. Bruckstein, D.~L. Donoho, and M.~Elad}, {\em From sparse solutions
  of systems of equations to sparse modeling of signals and images}, SIAM Rev.,
  51 (2009), pp.~34--81.

\bibitem{cai2009convergence}
{\sc J.~Cai, S.~Osher, and Z.~Shen}, {\em Convergence of the linearized bregman
  iteration for $\ell_1$-norm minimization}, Math. Comput., 78 (2009),
  pp.~2127--2136.

\bibitem{cai2009linearized}
\leavevmode\vrule height 2pt depth -1.6pt width 23pt, {\em Linearized bregman
  iterations for compressed sensing}, Math. Comput., 78 (2009), pp.~1515--1536.

\bibitem{Candes08}
{\sc E.~J. Cand{\`e}s}, {\em The restricted isometry property and its
  implications for compressed sensing}, C. R. Math. Acad. Sci. Paris, 346
  (2008), pp.~589--592.

\bibitem{CandesT05}
{\sc E.~J. Cand{\`e}s and T.~Tao}, {\em Decoding by linear programming}, IEEE
  Trans. Inform. Theory, 51 (2005), pp.~4203--4215.

\bibitem{Connolly1983Solventaccessible}
{\sc M.~L. Connolly}, {\em Solvent-accessible surfaces of proteins and nucleic
  acids}, Science, 221 (1983), pp.~709--713.

\bibitem{Cons_Wang_SIAM_2014}
{\sc P.~G Constantine, E.~Dow, and Q.~Wang}, {\em Active subspace methods in
  theory and practice: Applications to kriging surfaces}, SIAM J. Sci. Comput.,
  36 (2014), pp.~A1500--A1524.

\bibitem{DonohoET06}
{\sc D.~L. Donoho, M.~Elad, and V.~N. Temlyakov}, {\em Stable recovery of
  sparse overcomplete representations in the presence of noise}, IEEE Trans.
  Inform. Theory, 52 (2006), pp.~6--18.

\bibitem{DoostanO11}
{\sc A.~Doostan and H.~Owhadi}, {\em A non-adapted sparse approximation of
  {PDE}s with stochastic inputs}, J. Comput. Phys., 230 (2011), pp.~3015--3034.

\bibitem{Dror2012Biomolecular}
{\sc R.~O. Dror, R.~M. Dirks, J.~P. Grossman, H.~Xu, and D.~E. Shaw}, {\em
  Biomolecular simulation: A computational microscope for molecular biology},
  Annu. Rev. Biophys., 41 (2012), pp.~429--452.

\bibitem{FooK10}
{\sc J.~Foo and G.~E. Karniadakis}, {\em Multi-element probabilistic
  collocation method in high dimensions}, J. Comput. Phys., 229 (2010),
  pp.~1536--1557.

\bibitem{FooWK08}
{\sc J.~Foo, X.~Wan, and G.~E. Karniadakis}, {\em The multi-element
  probabilistic collocation method ({ME}-{PCM}): error analysis and
  applications}, J. Comput. Phys., 227 (2008), pp.~9572--9595.

\bibitem{GanapathyZ07}
{\sc B.~Ganapathysubramanian and N.~Zabaras}, {\em Sparse grid collocation
  schemes for stochastic natural convection problems}, J. Comput. Phys., 225
  (2007), pp.~652--685.

\bibitem{Genz_Kei_1996}
{\sc A.~Genz and B.D. Keister}, {\em Fully symmetric interpolatory rules for
  multiple integrals over infinite regions with gaussian weight}, J. Comput.
  Appl. Math., 71 (1996), pp.~299 -- 309.

\bibitem{GhanemS91}
{\sc R.~G. Ghanem and P.~D. Spanos}, {\em Stochastic finite elements: a
  spectral approach}, Springer-Verlag, New York, 1991.

\bibitem{Go_Noguti_PNAS_1983}
{\sc N.~Go, T.~Noguti, and T.~Nishikawa}, {\em Dynamics of a small globular
  protein in terms of low-frequency vibrational modes}, Proc. Natl. Acad. Sci.
  U.S.A., 80 (1983), pp.~3696--3700.

\bibitem{goldstein2009split}
{\sc T.~Goldstein and S.~Osher}, {\em {The split Bregman method for
  L1-regularized problems}}, SIAM J. Imaging Sci., 2 (2009), pp.~323--343.

\bibitem{GrantB}
{\sc M.~Grant and S.~Boyd}, {\em {CVX}: Matlab software for disciplined convex
  programming}.
\newblock \url{http://cvxr.com/cvx/}.

\bibitem{Hali_Bahar_1997}
{\sc T.~Haliloglu, I.~Bahar, and B.~Erman}, {\em Gaussian dynamics of folded
  proteins}, Phys. Rev. Lett., 79 (1997), pp.~3090--3093.

\bibitem{Harris2013Influence}
{\sc R.~C. Harris, A.~H. Boschitsch, and M.~O. Fenley}, {\em Influence of grid
  spacing in {Poisson–Boltzmann} equation binding energy estimation}, J.
  Chem. Theory Comput., 9 (2013), pp.~3677--3685.

\bibitem{Lee1971Interpretation}
{\sc B.~Lee and F.~M. Richards}, {\em The interpretation of protein structures:
  Estimation of static accessibility}, J. Mol. Biol., 55 (1971), pp.~379--IN4.

\bibitem{Levitt_Sander_JQC_1983}
{\sc M.~Levitt, C.~Sander, and P.~S. Stern}, {\em The normal modes of a
  protein: Native bovine pancreatic trypsin inhibitor}, Int. J. Quant. Chem.:
  Quantum Biology Symposium, 10 (1983), pp.~181--199.

\bibitem{Li10}
{\sc X.~Li}, {\em Finding deterministic solution from underdetermined equation:
  large-scale performance variability modeling of analog/rf circuits}, IEEE
  Trans. Comput.-Aided Design Integr. Circuits Syst., 29 (2010),
  pp.~1661--1668.

\bibitem{MaZ09}
{\sc X.~Ma and N.~Zabaras}, {\em An adaptive hierarchical sparse grid
  collocation algorithm for the solution of stochastic differential equations},
  J. Comput. Phys., 228 (2009), pp.~3084--3113.

\bibitem{MaZ10}
\leavevmode\vrule height 2pt depth -1.6pt width 23pt, {\em An adaptive
  high-dimensional stochastic model representation technique for the solution
  of stochastic partial differential equations}, J. Comput. Phys., 229 (2010),
  pp.~3884--3915.

\bibitem{McCammon_Harvey_1987}
{\sc A.~McCammon and S.~C. Harvey}, {\em Dynamics of protein and nucleic
  acids}, Cambridge University Press., Cambridge, 1987.

\bibitem{Nar_Xiu_Siam_2012}
{\sc A.~Narayan and D.~Xiu}, {\em Stochastic collocation methods on
  unstructured grids in high dimensions via interpolation}, SIAM J. Sci.
  Comput., 34 (2012), pp.~A1729--A1752.

\bibitem{Nar_Xiu_Siam_2013}
\leavevmode\vrule height 2pt depth -1.6pt width 23pt, {\em Constructing nested
  nodal sets for multivariate polynomial interpolation}, SIAM J. Sci. Comput.,
  35 (2013), pp.~A2293--A2315.

\bibitem{Nied92}
{\sc H.~Niederreiter}, {\em Random number generation and quasi-Monte Carlo
  methods}, SIAM, Philadelphia, PA, 1992.

\bibitem{NobileTW08}
{\sc F.~Nobile, R.~Tempone, and C.~G. Webster}, {\em An anisotropic sparse grid
  stochastic collocation method for partial differential equations with random
  input data}, SIAM J. Numer. Anal., 46 (2008), pp.~2411--2442.

\bibitem{petras_sg}
{\sc K.~Petras}, {\em Smolpack: a software for smolyak quadrature with
  clenshaw-curtis basis-sequence}.
\newblock
  \url{http://people.sc.fsu.edu/~jburkardt/c_src/smolpack/smolpack.html, 2003.}

\bibitem{Ponder2003Force}
{\sc J.~W. Ponder and D.~A. Case}, {\em Force Fields for Protein Simulations},
  vol.~66 of Advances in Protein Chemistry, Elsevier, 2003, pp.~27--85.

\bibitem{RauhutW12}
{\sc H.~Rauhut and R.~Ward}, {\em Sparse legendre expansions via
  $l_1$-minimization}, J. Approx. Theory, 164 (2012), pp.~517--533.

\bibitem{Ren2012Biomolecular}
{\sc P.~Ren, J.~Chun, D.~G. Thomas, M.~J. Schnieders, M.~Marucho, J.~Zhang, and
  N.~A. Baker}, {\em Biomolecular electrostatics and solvation: a computational
  perspective}, Q. Rev. Biophys., 45 (2012), pp.~427--491.

\bibitem{Richmond1984Solvent}
{\sc T.~J. Richmond}, {\em Solvent accessible surface area and excluded volume
  in proteins}, J. Mol. Biol., 178 (1984), pp.~63--89.

\bibitem{Najm_UQ_2012_a}
{\sc F.~Rizzi, H.~N. Najm, B.~J. Debusschere, K.~Sargsyan, M.~Salloum,
  H.~Adalsteinsson, and O.~M. Knio}, {\em Uncertainty quantification in md
  simulation. part i: Forward propagation}, Multiscale Model Simul., 10 (2012).

\bibitem{Najm_UQ_2012_b}
\leavevmode\vrule height 2pt depth -1.6pt width 23pt, {\em Uncertainty
  quantification in md simulation. part ii: Bayesian inference of force-field
  parameters}, Multiscale Model Simul., 10 (2012), pp.~1460--1492.

\bibitem{Roux1999Implicit}
{\sc B.~Roux and T.~Simonson}, {\em Implicit solvent models}, Biophys. Chem.,
  78 (1999), pp.~1--20.

\bibitem{Shrake_JMB_1973}
{\sc A.~Shrake and J.A. Rupley}, {\em Environment and exposure to solvent of
  protein atoms. lysozyme and insulin}, J. Mol. Biol., 79 (1973), pp.~351 --
  371.

\bibitem{Sloan94}
{\sc I.~H. Sloan and Joe S.}, {\em Lattice Methods for Multiple Integration},
  Oxford University Press, New York, 1994.

\bibitem{Tama_Sane_Protein_Eng_2001}
{\sc F.~Tama and Y.-H. Sanejouand}, {\em Conformational change of proteins
  arising from normal mode calculations}, Protein Eng., 14 (2001), pp.~1--6.

\bibitem{Tip_Ghanem_2014}
{\sc R.~Tipireddy and R.~Ghanem}, {\em Basis adaptation in homogeneous chaos
  spaces}, J. Comput. Phys., 259 (2014), pp.~304 -- 317.

\bibitem{Tirion_PRL_1996}
{\sc M.~M. Tirion}, {\em Large amplitude elastic motions in proteins from a
  single-parameter, atomic analysis}, Phys. Rev. Lett., 77 (1996),
  pp.~1905--1908.

\bibitem{BergF08}
{\sc E.~van~den Berg and M.~P. Friedlander}, {\em Probing the pareto frontier
  for basis pursuit solutions}, SIAM J. Sci. Comput., 31 (2008), pp.~890--912.

\bibitem{Wlodawer1984Structure}
{\sc A.~Wlodawer, J.~Walter, R.~Huber, and L.~Sj\"{o}lin}, {\em Structure of
  bovine pancreatic trypsin inhibitor. results of joint neutron and x-ray
  refinement of crystal form {II}.}, J. Mol. Biol., 180 (1984), pp.~301--329.

\bibitem{XiuH05}
{\sc D.~Xiu and J.~S. Hesthaven}, {\em High-order collocation methods for
  differential equations with random inputs}, SIAM J. Sci. Comput., 27 (2005),
  pp.~1118--1139.

\bibitem{XiuK02}
{\sc D.~Xiu and G.~E. Karniadakis}, {\em The {W}iener-{A}skey polynomial chaos
  for stochastic differential equations}, SIAM J. Sci. Comput., 24 (2002),
  pp.~619--644.

\bibitem{YanGX12}
{\sc L.~Yan, L.~Guo, and D.~Xiu}, {\em Stochastic collocation algorithms using
  $l_1$-minimization}, Int. J. Uncertainty Quantification, 2 (2012),
  pp.~279--293.

\bibitem{YangCLK12}
{\sc X.~Yang, M.~Choi, G.~Lin, and G.~E. Karniadakis}, {\em Adaptive {ANOVA}
  decomposition of stochastic incompressible and compressible flows}, J.
  Comput. Phys., 231 (2012), pp.~1587 -- 1614.

\bibitem{YangK13}
{\sc X.~Yang and G.~E. Karniadakis}, {\em Reweighted {$\ell_1$} minimization
  method for stochastic elliptic differential equations}, J. Comput. Phys., 248
  (2013), pp.~87 -- 108.

\bibitem{yin2008bregman}
{\sc W.~Yin, S.~Osher, D.~Goldfarb, and J.~Darbon}, {\em {B}regman iterative
  algorithms for $\ell_1$-minimization with applications to compressed
  sensing}, SIAM J. Imaging Sci., 1 (2008), pp.~143--168.

\bibitem{ZhangCK12}
{\sc Z.~Zhang, M.~Choi, and G.~E. Karniadakis}, {\em Error estimates for the
  {ANOVA} method with polynomial chaos interpolation: Tensor product
  functions}, SIAM J. Sci. Comput., 34 (2012), pp.~A1165--A1186.

\bibitem{ZhangTRK14}
{\sc Z.~Zhang, M.~V. Tretyakov, B.~Rozovskii, and G.~E. Karniadakis}, {\em A
  recursive sparse grid collocation method for differential equations with
  white noise}, SIAM J. Sci. Comput., 36 (2014), pp.~A1652--A1677.

\end{thebibliography}

\end{document}